\documentclass[aps,prapplied,reprint,fleqn]{revtex4-2}
\usepackage{amsmath}
\usepackage{bm}
\usepackage{amssymb}
\usepackage{graphicx}
\usepackage{xcolor}
\usepackage[utf8]{inputenc}
\usepackage[T1]{fontenc}
\usepackage{verbatim}
\usepackage[english]{babel}
\usepackage{blindtext}
\usepackage{float}
\usepackage{lipsum}
\usepackage{silence}
\usepackage{xparse}

\newcommand{\vE}{\mathbf{E}}
\newcommand{\vH}{\mathbf{H}}
\newcommand{\vEi}{\vE_{\parallel,\mathrm{i}}}
\newcommand{\vEt}{\vE_{\parallel,\mathrm{t}}}
\newcommand{\vHi}{\vH_{\parallel,\mathrm{i}}}
\newcommand{\vHt}{\vH_{\parallel,\mathrm{t}}}
\newcommand{\vHp}{\vH_{\parallel}}
\newcommand{\vEp}{\vE_{\parallel}}
\newcommand{\vEpav}{\vE_{\parallel,\mathrm{av}}}
\newcommand{\vHpav}{\vH_{\parallel,\mathrm{av}}}
\newcommand{\DvHp}{\Delta \mathbf{H}_{\parallel}}
\newcommand{\DvEp}{\Delta \mathbf{E}_{\parallel}}

\newcommand{\Yone}{\overline{\overline{Y}}_{1}}
\newcommand{\Ytwo}{\overline{\overline{Y}}_{2}}
\newcommand{\Ydel}{\overline{\overline{Y}}_{\Delta}}
\newcommand{\Ysig}{\overline{\overline{Y}}_{\Sigma}}

\newcommand{\Id}{\overline{\overline{I}}}

\newcommand{\Tmat}{\overline{\overline{T}}} % transmission matrix
\newcommand{\Rot}{\overline{\overline{\mathcal{R}}}_z}
\newcommand{\DeltaH}{\Delta\vHp}
\newcommand{\DeltaT}{\overline{\overline{\Delta}}_\mathrm{T}}
\newcommand{\AlphaT}{\overline{\overline{A}}_\mathrm{T}}

\newcommand{\Pp}{\mathbf{P}_\parallel}
\newcommand{\Mp}{\mathbf{M}_\parallel}
\newcommand{\Pz}{P_z}
\newcommand{\Mz}{M_z}

\newcommand{\kone}{\kappa_{1}}
\newcommand{\ktwo}{\kappa_{2}}
\newcommand{\Ezav}{E_\mathrm{z,avg}}
\newcommand{\Hzav}{H_\mathrm{z,avg}}

\newcommand{\zhat}{\mathbf{\hat{z}}}

\newcommand{\kpar}{\mathbf{k}_{\parallel}}

\newcommand{\erone}{\varepsilon_{\mathrm{r}1}}
\newcommand{\ertwo}{\varepsilon_{\mathrm{r}2}}

\newcommand{\murone}{\mu_{\mathrm{r}1}}
\newcommand{\murtwo}{\mu_{\mathrm{r}2}}

\newcommand{\epsz}{\varepsilon_{0}}

\newcommand{\Xee}{\overline{\overline{\chi}}_{\mathrm{ee}}}
\newcommand{\Xmm}{\overline{\overline{\chi}}_{\mathrm{mm}}}
\newcommand{\Xem}{\overline{\overline{\chi}}_{\mathrm{em}}}
\newcommand{\Xme}{\overline{\overline{\chi}}_{\mathrm{me}}}

\newcommand{\Xeepp}{\overline{\overline{\chi}}_{\mathrm{ee}}^{\parallel\parallel}}
\newcommand{\Xeepz}{\bm{\chi}_{\mathrm{ee}}^{\parallel z}}
\newcommand{\Xempp}{\overline{\overline{\chi}}_{\mathrm{em}}^{\parallel\parallel}}
\newcommand{\Xempz}{\bm{\chi}_{\mathrm{em}}^{\parallel z}}
\newcommand{\Xmmpp}{\overline{\overline{\chi}}_{\mathrm{mm}}^{\parallel\parallel}}
\newcommand{\Xmmpz}{\bm{\chi}_{\mathrm{mm}}^{\parallel z}}
\newcommand{\Xmepp}{\overline{\overline{\chi}}_{\mathrm{me}}^{\parallel\parallel}}
\newcommand{\Xmepz}{\bm{\chi}_{\mathrm{me}}^{\parallel z}}

\newcommand{\Xeezp}{\bm{\chi}_{\mathrm{ee}}^{z\parallel}}
\newcommand{\Xeezz}{\chi_{\mathrm{ee}}^{zz}}
\newcommand{\Xemzp}{\bm{\chi}_{\mathrm{em}}^{z\parallel}}
\newcommand{\Xemzz}{\chi_{\mathrm{em}}^{zz}}
\newcommand{\Xmmzp}{\bm{\chi}_{\mathrm{mm}}^{z\parallel}}
\newcommand{\Xmmzz}{\chi_{\mathrm{mm}}^{zz}}
\newcommand{\Xmezp}{\bm{\chi}_{\mathrm{me}}^{z\parallel}}
\newcommand{\Xmezz}{\chi_{\mathrm{me}}^{zz}}

\newcommand{\Xeepeff}{\overline{\overline{\chi}}_{\mathrm{ee}}^{\parallel,\mathrm{eff}}}
\newcommand{\Xempeff}{\overline{\overline{\chi}}_{\mathrm{em}}^{\parallel,\mathrm{eff}}}
\newcommand{\Xmmpeff}{\overline{\overline{\chi}}_{\mathrm{mm}}^{\parallel,\mathrm{eff}}}
\newcommand{\Xmepeff}{\overline{\overline{\chi}}_{\mathrm{me}}^{\parallel,\mathrm{eff}}}

\newcommand{\Xeezeff}{\bm{\chi}_{\mathrm{ee}}^{z,\mathrm{eff}}}
\newcommand{\Xemzeff}{\bm{\chi}_{\mathrm{em}}^{z,\mathrm{eff}}}
\newcommand{\Xmmzeff}{\bm{\chi}_{\mathrm{mm}}^{z,\mathrm{eff}}}
\newcommand{\Xmezeff}{\bm{\chi}_{\mathrm{me}}^{z,\mathrm{eff}}}

\newcommand{\Xiepeff}{\overline{\overline{\chi}}_{i\mathrm{e}}^{\parallel,\mathrm{eff}}}
\newcommand{\Ximpeff}{\overline{\overline{\chi}}_{i\mathrm{m}}^{\parallel,\mathrm{eff}}}
\newcommand{\Xiezeff}{\bm{\chi}_{i\mathrm{e}}^{z,\mathrm{eff}}}
\newcommand{\Ximzeff}{\bm{\chi}_{i\mathrm{m}}^{z,\mathrm{eff}}}

\newcommand{\Xiepp}{\overline{\overline{\chi}}_{i\mathrm{e}}^{\parallel\parallel}}
\newcommand{\Xiepz}{\bm{\chi}_{i\mathrm{e}}^{\parallel z}}
\newcommand{\Ximpp}{\overline{\overline{\chi}}_{i\mathrm{m}}^{\parallel\parallel}}
\newcommand{\Ximpz}{\bm{\chi}_{i\mathrm{m}}^{\parallel z}}

\newcommand{\tens}[3]{\ensuremath{#1_{\mathrm{#2}}^{#3}}}
\newcommand{\elem}[2]{\ensuremath{\chi_{\mathrm{#1}}^{#2}}}

\newcommand{\Q}{Q}                             % Q tens

\newcommand{\chiee}[1]{\elem{ee}{#1}}
\newcommand{\chime}[1]{\elem{me}{#1}}
\newcommand{\chiem}[1]{\elem{em}{#1}}
\newcommand{\chimm}[1]{\elem{mm}{#1}}

\newcommand{\Qee}[1]{\tens{\Q}{ee}{#1}}

\newcommand{\Smm}[1]{\tens{S}{mm}{#1}}

\begin{document}

\title{Generalized Invisibility in Metasurfaces}
%\title{Invisibility in Asymmetric Media via Bianisotropic Surface Responses}
\author{Mustafa Yücel}
\email{mustafa.yucel@epfl.ch}
\author{Karim Achouri}
\email{karim.achouri@epfl.ch}
\affiliation{
Institute of Electrical and Microengineering,
École Polytechnique Fédérale de Lausanne (EPFL),
Laboratory for Advanced Electromagnetics and Photonics,
Lausanne, Switzerland
}
%\date{\today}

\begin{abstract}

Electromagnetic invisibility, defined as reflectionless transmission with zero phase delay, imposes strict constraints on metasurface designs that go beyond conventional reflection suppression based on the Kerker effect. This condition can be viewed as a metasurface analogue of radiationless states such as anapole excitations. Here, we show that invisibility in metasurfaces embedded in identical media can only be achieved by introducing degrees of freedom, such as non-zero angle of incidence or higher-order multipolar responses. We demonstrate that, in dissimilar substrate and superstrate, achieving invisibility within a dipolar framework fundamentally requires pure bianisotropic coupling, while purely electric and magnetic responses are insufficient for lossless, passive and reciprocal systems. Using effective surface susceptibilities that account for the surrounding media and transverse wave vector, we derive closed-form conditions for both co- and cross-polarized invisibility. Importantly, we also demonstrate that the required bianisotropy does not need to be intrinsic, as an effective bianisotropic response may be achieved with anisotropic metasurface in dissimilar media leading to magnetoelectric coupling. Full-wave simulations of a metasurface at an air-dielectric interface confirm invisibility under oblique incidence. This work establishes a universal dipolar framework for invisible meta-optics in practically realistic scenarios. 

\end{abstract}

\maketitle
% \section{Introduction}
% \label{sec:introduction}

\section{Introduction}
\label{sec:introduction}

\begin{figure*}[t]
    \centering
    \includegraphics[width=0.8\linewidth]{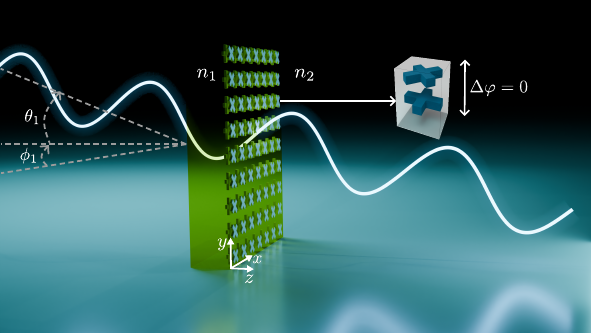}
    \caption{Conceptual illustration of electromagnetic invisibility through a metasurface. A structured metasurface (gray slab with subwavelength inclusions) is illuminated by an obliquely incident electromagnetic wave. 
    Despite the presence of the scatterers, the transmitted electric and magnetic fields (red and blue curves) emerge with unchanged amplitude, phase, and propagation direction, indicating reflectionless transmission and zero additional phase delay. The continuity of the wavefront across the metasurface highlights the invisibility condition derived in this work.}
    \label{fig:placeholder}
\end{figure*}

Metasurfaces provide a versatile platform for controlling electromagnetic waves using planar arrays of subwavelength scatterers~\cite{decker2015,chen2016a,li2018metasurfaces,Epstein2016,achouri2021}. By engineering their geometry and material composition, metasurfaces enable precise manipulation of amplitude, phase, polarization, and wavefront shaping~\cite{veksler2015multiple,balthasar2017metasurface,kildishev2013planar,huang2023leaky,yucel2025angle}, with applications ranging from quantum photonics to imaging and display technologies~\cite{solntsev2021metasurfaces,li2017nonlinear,dorrah2021metasurface,shi2025flat,gopakumar2024full,shi2024electrical}.

Among these functionalities, \textit{electromagnetic invisibility} represents a particularly interesting and fundamental concept, which is here defined as reflectionless scattering with zero-phase delay. This concept of invisibility was first mentioned by Robert W. Wood in 1902 who described how transparent objects become invisible when immersed in a medium of the same refractive index~\cite{wood1902invisibility}. Much later, Devaney and Wolf introduced several theorems related to non-radiative source theory that may be used to theoretically understand invisibility~\cite{devaney1973radiating}. Later, Kerker theoretically demonstrated that non-absorbing ellipsoids composed of inner and outer dielectric shells could become invisible provided the right choice of refractive indices~\cite{kerker1975invisible}. More recently, invisibility has been described and achieved by exploiting the concept of an anapole state, which corresponds to a destructive interference between a given primitive multipole moment and its corresponding toroidal moments, leading to a non-radiating excitation that renders lossless single or bulk particles as well as metasurfaces invisible in the far field~\cite{miroshnichenko2015nonradiating,yang2019nonradiating,basharin2023selective,hassan2025anapole,bukharin2026total}.

Note that the concept of an invisible metasurface differs from the concept of optical cloaking, where an external structure is used to conceal an object~\cite{alu2005achieving,ni2015ultrathin,monticone2016invisibility,labate2017surface,xu2021polarization}. Here, the metasurface itself is non-radiating while being potentially able to perform other desired operations. It should be noted that, while reflection suppression can be achieved in Huygens metasurfaces via the Kerker effect through balanced electric and magnetic dipolar responses~\cite{decker2015,Epstein2016,liu2018b,achouri2021}, enforcing zero transmission phase simultaneously imposes stronger constraints that cannot be satisfied by conventional dipolar metasurface designs due to limited degrees of freedom, as we shall soon demonstrate. As a result, non-trivial invisibility cannot be achieved in a passive dipolar metasurface placed in a homogeneous background medium, and previously reported realizations essentially rely on higher-order multipolar interference~\cite{terekhov2019,zhang2021wide,tuz2021,kuznetsov2021,poleva2023}. Moreover, the conditions for metasurface invisibility under arbitrary incidence and in the presence of different substrate and superstrate, which is the most practical scenario for a metasurface operating in the optical regime, remains unexplored.

The main objective of this work is thus to develop a general explanation of the concept of invisible scattering in metasurfaces. To do so, we provide rigorous conditions, given in terms of effective material parameters, to achieve invisibility in the case of a metasurface placed between different substrate and superstrate. For this purpose, we use a metasurface modeling framework based on the generalized sheet transitions conditions (GSTC)~\cite{kuester2003averaged,achouri2021}, which uses effective surface susceptibilities to capture the electromagnetic response of the metasurface. Due to the complexity of including multipolar effective material parameters~\cite{achouri2022b,tiukuvaara2023}, we shall restrict our analysis to dipolar responses. For completeness, we will nonetheless briefly discuss how the addition of multipolar contributions affect our results.

Our developments show that, in the case of a reciprocal \emph{dipolar} metasurface embedded in a homogeneous background medium, invisibility may be achieved only at oblique incidence. This happens because the transverse component of the incident wavevector introduces an additional degree of freedom that couples tangential and normal fields, leading to invisibility. We further show that this type of invisibility arises from a destructive interference between a tangential electric dipole and a normal magnetic dipole (or vice versa), a mechanism that cannot be achieved at normal incidence. This can be viewed as an oblique-incidence extension of the conventional Kerker condition that leads to a simultaneous cancellation of backward scattering and zero-phase delay transmission. We then analyze the general case of asymmetric background media and show that co- and cross-polarized invisibility can be achieved in dipolar metasurfaces and that it requires purely magnetoelectric responses. Moreover, we also demonstrate that the required magnetoelectric coupling need not be intrinsic to the metasurface, but may instead arise from environmental asymmetry. These developments are validated through full-wave simulations of metasurfaces at dielectric interfaces that demonstrate invisible scattering in practically realistic scenarios. 

% Note that while multilayer non-scattering implementations exist, they do not address the limits of single-layer dipolar metasurfaces~\cite{cuesta2019nonscattering}.

\section{Problem formulation}
\label{sec:problem}
We consider an electrically thin metasurface located at $z=0$ separating two homogeneous media. The metasurface is spatially uniform and is composed of a single deeply sub-wavelength unit cell that is periodically repeated in the $xy$-plane. 
Region~1 ($z<0$) is characterized by $(\epsilon_1,\mu_1)$ with wave impedance $\eta_1$ and wavenumber $k_1$, while region~2 ($z>0$) is described by $(\epsilon_2,\mu_2)$ with $\eta_2$ and $k_2$.

A monochromatic plane wave of angular frequency $\omega$ impinges from region~1 at an incidence angle $\theta_1$ in the $xz$-plane, leading to the tangential wavevector being $\mathbf{k}_\parallel = k_1\sin\theta_1\,\hat{\mathbf{x}}$. Due to the translational invariance of the metasurface, the tangential component of the wavevector is conserved across the interface, $\mathbf{k}_{\parallel,1}=\mathbf{k}_{\parallel,2}$, which leads to Snell's law $n_1\sin\theta_1=n_2\sin\theta_2$.

Throughout this work, TE polarization denotes an electric field perpendicular to the plane of incidence ($\mathbf{E}\parallel\hat{\mathbf{y}}$), whereas TM polarization denotes a magnetic field perpendicular to the plane of incidence ($\mathbf{H}\parallel\hat{\mathbf{y}}$). 
The tangential field components at the metasurface are used in the following sections to define the field averages and discontinuities entering the GSTC formulation.

Our definition of invisibility requires no added transmission phase and full power transmission through the metasurface. For a lossless metasurface separating two dissimilar media, power conservation requires the transmitted field amplitude to include a normalization factor accounting for the difference in wave impedance and propagation angle. The corresponding transmission amplitude is therefore
\begin{equation}
\label{eq_tau}
\tau = \sqrt{\frac{n_1\cos\theta_1}{n_2\cos\theta_2}},
\end{equation}
which ensures that the normal component of the Poynting vector is preserved across the interface.
\section{GSTC Formulation}

The mathematical foundation of this work is based on the GSTC, which describes the boundary conditions of a metasurface based on the presence of polarization densities~\cite{kuester2003averaged}. Considering only dipolar responses, the GSTC are expressed as
\begin{subequations} \label{gstc}
\begin{align}
    \DvHp &= - j\omega\zhat \times \Pp - \frac{1}{ \mu_0} \nabla_{\parallel} M_z, \label{gstc_a}\\
    \DvEp &= + j\omega\zhat \times \Mp - \frac{1}{ \epsilon_0} \nabla_{\parallel} P_z \label{gstc_b},
\end{align}
\end{subequations}
where $\DvEp$ and $\DvHp$ are the tangential electric and magnetic field differences at the interface, $\Pp$ and $\Mp$ are the tangential electric and magnetic polarization densities, and $P_z$ and $M_z$ the electric and magnetic components of the polarization densities normal to the metasurface. 

The polarization densities are related to the averaged fields $\mathbf{E}_\text{av}$ 
and $\mathbf{H}_\text{av}$ and to a full bianisotropic susceptibility tensor, including both tangential and normal electric and magnetic components. The complete constitutive relations are provided in the SI.

Note that imposing invisibility requires that the reflected fields be zero, i.e., $\mathbf{E}_\text{r} = \mathbf{H}_\text{r} = 0$, which affects the definition of both the field differences and averages.  

\section{Review of invisibility in a symmetric environment}
\subsection{Reflectionless Huygens Metasurface}

A Huygens metasurface achieves reflection cancellation when the electric and magnetic dipolar responses radiate fields that destructively interfere in the backward direction. This response is achievable by engineering the metasurface to obtain the desired tangential susceptibilities that satisfy the Kerker condition~\cite{liu2018b}. However, such a reflectionless metasurface is not necessarily invisible at some frequencies according to our definition. In this context, the reflectionless scattering is achieved if and only if the electric and magnetic susceptibilities are compensating each other such that the reflection coefficient is \textit{zero}. For instance, for a TM-polarized normally incident plane wave impinging on an isotropic metasurface, the reflectionless condition that applies to the susceptibilities is simply $\chiee{xx} = \chimm{yy}$~\cite{achouri2021}. With this susceptibility condition, the metasurface transmission coefficient becomes unity in amplitude, but the phase remains non-zero and equal to (see SI)
\begin{align}\label{huygen_t_phase}
    \angle T = \arctan \left [ \frac{4k\chiee{xx}}{\left ( k \chiee{xx} \right )^2-4} \right].
\end{align}
It follows that the only possibility to achieve invisibility with such a metasurface is by having $\chiee{xx} = \chimm{yy} = 0$. This is achievable either if there is no metasurface at all (trivial case) or if the Lorentzian response of each susceptibility crosses at \textit{zero}, which might be possible by using low-loss resonators that exhibit multiple Lorentzian resonances.
\subsection{Limitations of dipolar invisibility in \\ symmetric media}
Using the GSTC~\eqref{gstc} for a symmetric environment ($n_1 = n_2$) and imposing the invisibility condition ($\mathbf{E}_\text{t} = \mathbf{E}_\text{i}$ and $\mathbf{H}_\text{t} = \mathbf{H}_\text{i}$), the tangential field discontinuities vanish, $\DvEp = \DvHp = 0$, while the induced polarizations remain nonzero ($\mathbf{P} \neq 0$, $\mathbf{M} \neq 0$). Substituting these conditions into~\eqref{gstc} yields
\begin{subequations} \label{gstc_inv_same_media_PM}
\begin{align}
    \omega \mu_0 \Pp       &=  - \zhat \times\kpar M_z, \label{gstc_inv_same_media_PM_a}\\
    \omega \epsilon_0 \Mp  &=  + \zhat \times\kpar P_z \label{gstc_inv_same_media_PM_b},
\end{align}
\end{subequations}
where the mapping $\nabla_\parallel \rightarrow -j\mathbf{k}_\parallel$ has been used for plane-wave excitation ($e^{-j\kpar\cdot\mathbf{r}}$).

It was reported in~\cite{terekhov2019,zhang2021wide} that invisibility can be achieved at normal incidence, which can be directly inferred from~\eqref{gstc_inv_same_media_PM}. At normal incidence, the right-hand sides of~\eqref{gstc_inv_same_media_PM} vanish because $\kpar = 0$. Consequently, we must have $\Pp = \Mp = 0$, which corresponds to a complete suppression of dipolar radiation. This can be interpreted as a double anapole state~\cite{terekhov2019,zhang2021wide,basharin2023selective}.

Introducing the susceptibilities and noting that, for identical media, the average fields reduce to the incident fields (i.e., $\mathbf{E_\mathrm{av}} = \mathbf{E_\mathrm{i}}$ and $\mathbf{H_\mathrm{av}} = \mathbf{H_\mathrm{i}}$), these conditions yield, for a TM-polarized normally incident wave, $\chi_\mathrm{ee}^{xx} = -\chi_\mathrm{em}^{xy}$ and $\quad \chi_\mathrm{mm}^{yy} = -\chi_\mathrm{me}^{yx}.$
Imposing reciprocity, $\Xme = -\Xem^T$, where the superscript $T$ denotes the transpose operation, leads to the invisibility condition
\begin{equation}
\label{eq_cond_norm_biani}
    \chi_\text{ee}^{xx} = -\chi_\text{mm}^{yy} = -\chi_\text{em}^{xy}.
\end{equation}
This condition fundamentally differs from the conventional Kerker condition, which requires a balance of tangential electric and magnetic dipoles. Here, invisibility requires a precise relation between electric, magnetic, and magnetoelectric responses.

Equation~\eqref{eq_cond_norm_biani} admits two possible solutions. The first corresponds to the trivial case where all susceptibilities vanish, $\chi_\mathrm{ee}^{xx} = \chi_\mathrm{mm}^{yy} = \chi_\mathrm{em}^{xy} = 0$. This situation may be realized for a symmetric system, which enforces $\chi_\mathrm{em}^{xy}=0$~\cite{achouri2023spatial}, and by tuning the electric and magnetic responses to simultaneously cross zero at a given frequency $\omega_L$, e.g., through double Lorentzian resonances.

The second possibility corresponds to nonzero susceptibilities satisfying~\eqref{eq_cond_norm_biani}. However, such a solution cannot be achieved in a passive and lossless metasurface. Indeed, the Poynting theorem imposes that $\chi_\mathrm{ee}^{xx}$ and $\chi_\mathrm{mm}^{yy}$ are real-valued, whereas $\chi_\mathrm{em}^{xy}$ must be purely imaginary~\cite{achouri2021}. This fundamental incompatibility prevents the condition~\eqref{eq_cond_norm_biani} from being satisfied in a passive and lossless metasurface.

Therefore, dipolar metasurfaces cannot achieve non-trivial invisibility at normal incidence in symmetric environments. This limitation motivates the exploration of alternative mechanisms, such as higher-order multipolar responses~\cite{terekhov2019,zhang2021wide} or oblique-incidence configurations, as discussed in the following.
\subsection{Quadrupolar invisibility}
\label{sec_quad}

We have seen that dipolar metasurfaces are somewhat limited in their capabilities to achieve invisibility. We now show, based on the suggestions in~\cite{terekhov2019,zhang2021wide}, that the introduction of quadrupolar responses in the GSTC formalism helps overcome this limitation. 

For simplicity, we consider the case of a metasurface that is surrounded by two identical media and exhibits both electric and magnetic dipolar and quadrupolar contributions. Assuming a normally incident plane wave, the GSTC may be rewritten as~\cite{allahverdizadeh2025multipolar}
\begin{subequations}\label{gstc_multipolar}
\begin{align}
    j\omega  \Mp
    &= + \frac{k^2}{2}\,\zhat \times \left ( \overline{\overline{Q}} \cdot \zhat \right ), \label{gstc_multipolar:a}\\
    j\omega \Pp
    &= - \frac{k^2}{2}\,\zhat \times \left ( \overline{\overline{S}} \cdot \zhat \right ), \label{gstc_multipolar:b}
\end{align}
\end{subequations}
where $\overline{\overline{Q}}$ and $\overline{\overline{S}}$ denote the electric and magnetic quadrupolar tensors, respectively. Assuming a symmetric metasurface, we neglect the presence of magnetoelectric coupling susceptibilities~\cite{achouri2023spatial}, this reduces the multipolar surface polarization densities to~\cite{tiukuvaara2023}
\begin{subequations}\label{multipolar_surface_densities}
\begin{align}
P_i   &= \varepsilon_0\, \chi_\mathrm{ee}^{ij}\, E_{\mathrm{av},j}, \label{multipolar_surface_densities_a}\\
M_i   &= \mu_0\chi_\mathrm{mm}^{ij}\, H_{\mathrm{av},j}, \label{multipolar_surface_densities_b}\\
Q_{il} &= \frac{c}{2k^{2}\varepsilon_0}\, \Qee{\prime\, iljk}\, \partial_z E_{\mathrm{av},j}, \label{multipolar_surface_densities_c}\\
S_{il} &= \frac{1}{2k^{2}}\, \Smm{\prime\, iljk}\, \partial_z H_{\mathrm{av},j}. \label{multipolar_surface_densities_d}
\end{align}
\end{subequations}
Combining~\eqref{gstc_multipolar} and~\eqref{multipolar_surface_densities}, and considering TM-polarized waves, we can find a condition of invisibility on the dipolar and quadrupolar susceptibilities given by
\begin{subequations}
\label{gstc_multipolar_invisibility_condition}
\begin{align}
    4 \, \chi_\mathrm{ee}^{xx} &= -\Smm{\prime \, yzzy}, \\
    4 \, \chi_\mathrm{mm}^{yy} &= -\Qee{\prime \, xzzx}.
\end{align}
\end{subequations}

This result is perfectly consistent with the one found in~\eqref{eq_cond_norm_biani} since, in the absence of quadrupolar susceptibilities and for a symmetric metasurface, i.e. $\Smm{\prime \, yzzy}=\Qee{\prime \, xzzx}=0$ and $\chiem{xy}=0$, we retrieve the condition that $\chiee{xx}=0$ and $\chimm{yy}=0$ to achieve invisibility. Note that the conditions in \eqref{gstc_multipolar_invisibility_condition} explicitly validate the findings in~\cite{zhang2021wide}, showing that, to achieve invisibility, the electric dipole must be coupled to the magnetic quadrupole, and the magnetic dipole to the electric quadrupole. Importantly, Eq.~\eqref{gstc_multipolar_invisibility_condition} shows that, by including multipolar responses, we can break the constraints of~\eqref{eq_cond_norm_biani} that required either a combination of gain and loss or that all susceptibilities be zero. Instead, we may achieve invisibility by properly balancing dipolar and quadrupolar responses in a gainless and lossless fashion.

We have seen that restrictions with dipolar GSTC modeling may be alleviated when introducing higher-order multipolar contributions. Thus, in the remainder of this paper, we concentrate our attention purely on dipolar cases, for simplicity and convenience. Keeping in mind that dipolar responses that seem impossible to achieve might be achievable within a multipolar modeling framework. 

\subsection{Dipolar and oblique incidence}

The previous analysis showed that dipolar metasurfaces are fundamentally constrained at normal incidence in symmetric environments. We now demonstrate that these limitations can be lifted under oblique illumination.
To illustrate this mechanism, we consider plane-wave incidence in the $xz$-plane on an isotropic metasurface. In this configuration, invisibility can be achieved through the interaction between a tangential electric polarization component $P_y$ and a normal magnetic polarization component $M_z$, as follows from~\eqref{gstc_inv_same_media_PM_a}. Assuming that only these two dipolar responses are present, the invisibility condition reduces to
\begin{align}
\label{eq_cond_PyMz}
P_y = - \frac{M_z}{\eta}\sin(\theta),
\end{align}
where $\theta$ is the incidence angle and $\eta$ is the wave impedance of the surrounding medium. 

Substituting the constitutive relations~\eqref{multipolar_surface_densities_a} and~\eqref{multipolar_surface_densities_b} into~\eqref{eq_cond_PyMz} yields
\begin{equation}
\label{eq_cond_PyMz_X}
    \chiee{yy} = -\chimm{zz}\sin{\theta}^2.
\end{equation}

This result reveals a fundamentally different mechanism from the conventional Kerker condition, which relies on a balance between tangential electric and magnetic dipoles. In Eq.~\eqref{eq_cond_PyMz_X}, reflection cancellation arises from the destructive interference between a tangential electric dipole and a \emph{normal} magnetic dipole.

The cancellation condition is therefore angle dependent, and perfect invisibility is achieved only at a specific incidence angle. Nevertheless, it demonstrates that invisibility can be achieved with \emph{only} dipolar responses in a reciprocal, lossless and gainless system; something that is not possible to achieve at normal incidence, as we have seen in Sec.~\ref{sec_quad}.

This result highlights the importance of the normal polarizations under oblique illumination and provides a key insight for overcoming the limitations of dipolar metasurfaces without resorting to higher-order multipolar responses.

\section{General Synthesis Equations}

In this section, we analyze the general conditions derived in the SI for electromagnetic invisibility in metasurfaces under arbitrary incidence angle, polarization and surrounding media. Building upon the GSTC framework, we express the metasurface response directly in terms of effective susceptibilities that incorporate both the intrinsic bianisotropic properties and the influence of the embedding media. This leads to the general synthesis equations for invisibility given by
\begin{subequations}
\label{general1}
\begin{equation}
\begin{split}
\pm \Rot\cdot &\Ydel \cdot \vEi
= \Big[
- j \omega \Rot\cdot \left( \epsz \Xeepeff + \frac{1}{c_0} \Xempeff \right) \\
&\quad + j \frac{\kpar}{\mu_0} \left( \mu_0 \Xemzeff + \frac{1}{c_0} \Xmezeff \right)
\Big] \cdot \vEi,
\end{split}
\end{equation}
\begin{equation}
\label{general2}
\begin{split}
\DeltaT \cdot &\vEi
= 
\Big[
+ j \omega\Rot\cdot 
\big( \mu_0 \Xmmpeff 
+ \frac{1}{c_0} \Xmepeff \big) \\
&\quad +j\frac{\kpar}{\epsz} 
\big( \epsz \Xeezeff + \frac{1}{c_0} \Xemzeff \big)
\Big] \cdot \vEi,
\end{split}
\end{equation}
\end{subequations}
%

 % $\AlphaT$ and $\AlphaT$ are effective transfer dyadic, $\Ysig$ and $\Ysig$ are sum and weighted admittance tensors

where $\DeltaT$, $\Ydel$ and $\Rot$ are the differentiated transfer dyadic, admittance tensor and the $z$-oriented rotation matrix, respectively, which are defined in the SI. The superscript ``eff'' denotes effective susceptibilities that incorporate the intrinsic bianisotropic response of the metasurface, the transverse wavevector, and the surrounding media through the averaged-field relations. The incident electric field $\vEi$ is kept to properly define the polarization of the incident field. These relations provide a general dipolar framework for engineering invisible metasurfaces and thus represent the main result of this paper entirely derived in the SI.

% These equations in~\eqref{general1} reveal three key physical mechanisms. First, the presence of a nonzero transverse wavevector couples tangential and normal polarization components. Second, the asymmetry between the surrounding media modifies the effective response of the metasurface. Third, these effects combine to produce an effective bianisotropic behavior, even when the intrinsic structure is purely anisotropic.

Each effective susceptibility contains contributions from both tangential field interactions ($\parallel$) and normal polarizations ($z$) modulated by the transverse wavevector $\kpar$, introducing additional degrees of freedom not present at normal incidence. The latter vanish at normal incidence ($\mathbf{k}_\parallel=0$), which explains why normal polarizations become essential for achieving invisibility under oblique illumination. As a consequence, the effective susceptibilities are functions of the intrinsic susceptibilities, the regions $n_1$ and $n_2$ and the tangential wavevector,
\begin{equation}
\overline{\overline{\chi}}_{ij}^{\mathrm{eff}}
= f\!\left(\overline{\overline{\chi}}_{ij}, \kpar, n_1, n_2\right),
\qquad i,j \in \{\mathrm{e,m}\}.
\end{equation}
Despite being effective, the nature of the effective susceptibilities remains unchanged compared to the original ones (i.e., $\overline{\overline{\chi}}_\text{ee}^\text{eff}$ depends only on the components of $\Xee$, and so on).

Equations~\eqref{general1} are fully general and apply to metasurfaces that may exhibit nonreciprocity as well as loss or gain. In practice, for a passive and lossless system, the electric and magnetic susceptibilities $\Xee$ and $\Xmm$ are real-valued, while the magnetoelectric susceptibilities $\Xem$ and $\Xme$ are purely imaginary. In addition, reciprocity requires that $\Xme = -\Xem^T$~\eqref{eq_recip}. Applying these constraints to the relations in~\eqref{general1} leads to the reduced system given in~\eqref{eq_general_lossless} in the SI. It results that the invisibility conditions depend exclusively on the magnetoelectric coupling tensor $\Xem$ (and $\Xme$ by reciprocity), while the purely electric and magnetic responses must vanish, i.e., $\Xee = \Xmm = 0$. This result demonstrates that bianisotropy is not merely advantageous but fundamentally required to achieve invisibility under oblique incidence in asymmetric environments.

From a symmetry perspective~\cite{poleva2023,achouri2023spatial}, the emergence of magnetoelectric coupling requires breaking inversion symmetry. While such symmetry breaking is typically introduced through intrinsically bianisotropic unit cells and hard to realize in nanofabrication, we show that it can also arise at the system level. Specifically, an intrinsically \emph{anisotropic} metasurface embedded in dissimilar surrounding media exhibits an effective bianisotropic response due to the \emph{asymmetry} of the environment, leading to magnetoelectric coupling.
\begin{figure}[H]
    \centering
    \includegraphics[width=1\linewidth]{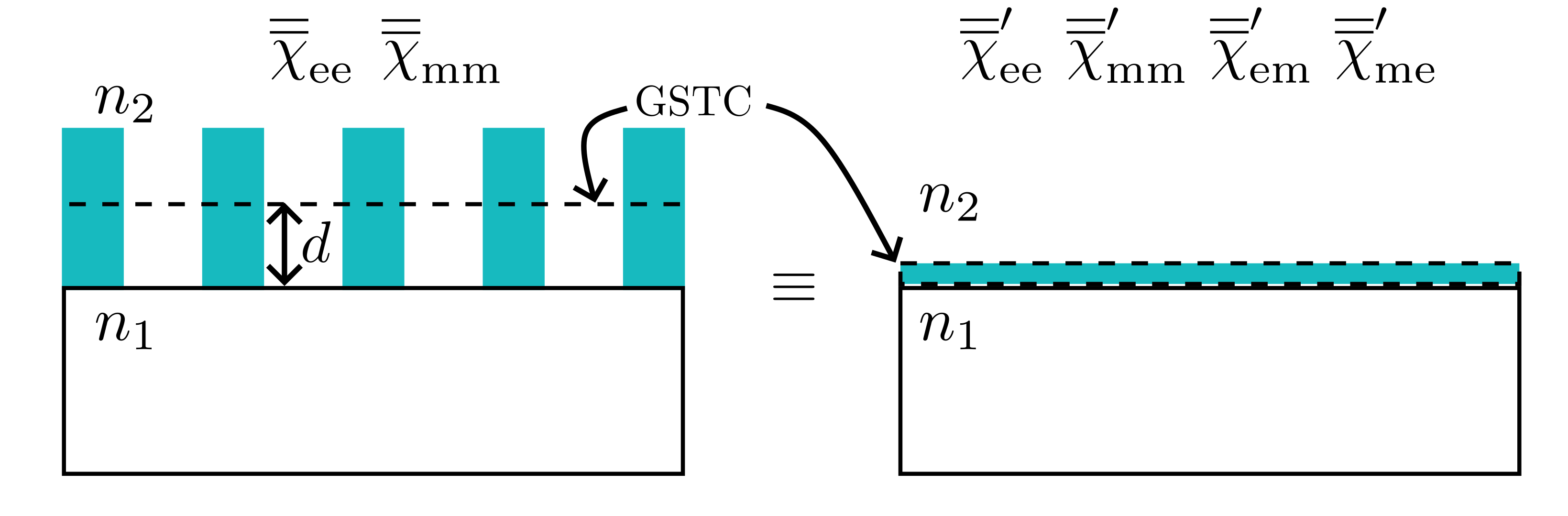}
    \caption{A thick anisotropic metasurface may be modeled as an equivalent zero-thickness bianisotropic metasurface.}
    \label{ms_equivalence_main}
\end{figure}
An illustration of this concept is depicted in Fig.~\ref{ms_equivalence_main}, where an anisotropic metasurface made of an array of scattering particles of height $2d$, and mirror symmetric along the normal direction, is placed at the interface between media $n_1$ and $n_2$. Since the metasurface is anisotropic, it is only modeled with the susceptibility tensors $\Xee$ and $\Xmm$. To trigger an effective bianisotropic response from this structure, we place the corresponding GSTC boundary at the center of mass of the scatterers, i.e., at a height $d$ above the substrate. 

Since we have the freedom to arbitrarily choose the position of the GSTC boundary without affecting the overall scattering response of the metasurface~\cite{allahverdizadeh2025multipolar}, we now consider an alternative situation where the GSTC is placed right at the interface between the two media, as shown on the right-side of Fig.~\ref{ms_equivalence_main}. In this case, the scattering response of the metasurface cannot be modeled with only $\Xee$ and $\Xmm$ and should rather be modeled as a fully bianisotropic structure with the effective susceptibilities $\Xee'$, $\Xmm'$, $\Xem'$ and $\Xme'$  all being functions of $d$. The height $d$ can thus be used to, for instance, modulate the ``effective'' susceptibility \chiem{xy\prime} proportionally to $\sin(kd)$ as described in~\eqref{effective_chi_substrate} of the SI. This methodology allows us to achieve the invisibility conditions in~\eqref{eq_general_lossless} in a nano-fabrication friendly way, as it lowers the number of fabrication steps and avoids complicated alignment challenges.

\begin{figure*}[t]
    \centering
    \includegraphics[width=\linewidth]{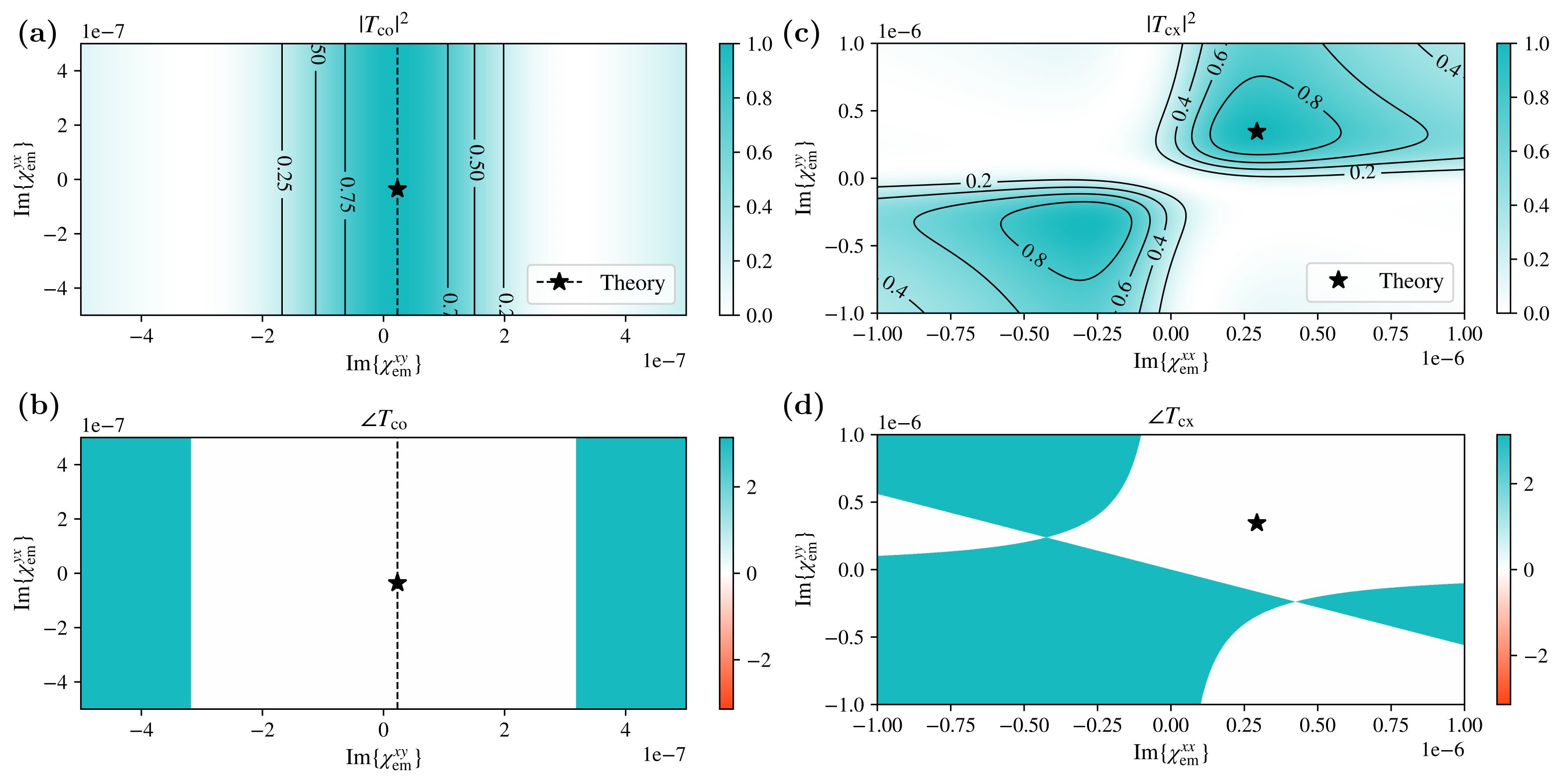}
    \caption{Power and phase maps for co- and cross-polarized invisibility for varying values of the susceptibilities around the optimal invisibility points indicated by the stars. \textbf{(a)},\textbf{(c)} Reflectance spectrum of the co- and cross-polarized transmission coefficient $T_{\mathrm{co}}$ and $T_{\mathrm{cx}}$ under TM incidence at $\theta_i = 30^\circ$ for an interface between air ($n_1=1$) and $\mathrm{SiO}_2$ ($n_2=1.45$). \textbf{(b)},\textbf{(d)} Corresponding transmission phase spectrum for both co- and cross- polarization, respectively.}
    \label{fig_Tcx}
\end{figure*}

\section{Examples of Metasurface Invisibility}

In this section, we illustrate the application of the general invisibility condition~\eqref{general1} with two examples, one for co-polarized and and one for cross-polarized scattering. For simplicity, we only consider scattering within the $xz$-plane for both TE and TM waves and include the presence of different media on both sides of the metasurface.

\subsection{Co-Polarization Invisibility}

For co-polarized invisibility, the transmitted field preserves the polarization state of the incident wave while eliminating reflection. To enforce this condition, we restrict the metasurface response to susceptibility components that do not mix orthogonal polarizations. In particular, we consider the magnetoelectric terms $\chi_{\mathrm{em}}^{xy}$ and $\chi_{\mathrm{em}}^{yx}$ together with their reciprocal counterparts, while other cross-polarizing components are set to zero.

Substituting these susceptibilities into the reduced GSTC relations~\eqref{eq_general_lossless} yields the condition for co-polarized invisibility, for both TE and TM incidence,

\begin{subequations}
\label{copol}
\begin{align}
\chi_\mathrm{em}^{yx} = -\chi_\mathrm{me}^{xy}
= \frac{2j}{k}\left(\frac{\tau - 1}{\tau + 1}\right),\\
\chi_\mathrm{em}^{xy} = -\chi_\mathrm{me}^{yx}
= \frac{2j}{k}\left(\frac{\tau - \Gamma}{\tau + \Gamma}\right),
\end{align}
\end{subequations}
where $\Gamma = n_1/n_2$ is the medium ratio and $\tau$ is given in~\eqref{eq_tau}. Since $\tau$ is angle dependent it follows that, when these conditions are satisfied, the metasurface is fully invisible at a given incidence angle and wavelength.

To demonstrate the validity of relations~\eqref{copol}, we plot in Fig.~\ref{fig_Tcx}(a) and Fig.~\ref{fig_Tcx}(b) the co-polarized transmittance and transmission phase, respectively, assuming a TM-polarized obliquely propagating incident wave at $\theta = 30^\circ$, for varying values of $\chiem{xy}$ and $\chiem{yx}$. In Fig.~\ref{fig_Tcx}(a), we see that at a specific value of $\chiem{xy}$, the co-polarized transmission is $|T_\text{co}|^2=1$, and the phase remains 0. The dashed line with the star represents the invisibility region described in~\eqref{copol}.

\subsection{Cross-Polarization Invisibility}

For cross-polarized invisibility, the transmitted field swaps its polarization with respect to the incident one and eliminates any reflection. Similarly to the previous case, we analyze the susceptibilities that mixes polarization components once the electromagnetic wave interacts with the metasurface. In particular, we consider the diagonal magnetoelectric components 
$\chi_{\mathrm{em}}^{xx}$ and $\chi_{\mathrm{em}}^{yy}$, together with their reciprocal counterparts $\chi_{\mathrm{me}}^{xx}$ and $\chi_{\mathrm{me}}^{yy}$.

Substituting these susceptibilities into the reduced invisibility GSTC equations~\eqref{eq_general_lossless} yields the conditions for cross-polarized invisibility. In the case of a TE-polarized incident wave converted into TM polarization, the required magnetoelectric susceptibilities are
\begin{subequations}
\label{eq_TM_to_TE}
\begin{align}
\chi_{\mathrm{em}}^{xx} &= -\chi_{\mathrm{me}}^{xx}
= -\,\frac{2j}{k}\,\frac{1}{\tau\,\cos\theta_2}, \\
\chi_{\mathrm{em}}^{yy} &= -\chi_{\mathrm{me}}^{yy}
= -\,\frac{2j}{k}\,\tau\,\cos\theta_2 .
\end{align}    
\end{subequations}
Conversely, for a TM-polarized incident wave converted into TE polarization, the susceptibilities become
\begin{subequations}
\label{eq_TE_to_TM}
\begin{align}
\chi_{\mathrm{em}}^{xx} &= -\chi_{\mathrm{me}}^{xx}
= +\,\frac{2j}{k}\,\frac{\tau}{\cos\theta_1}, \\
\chi_{\mathrm{em}}^{yy} &= -\chi_{\mathrm{me}}^{yy}
= +\,\frac{2j}{k}\,\frac{\cos\theta_1}{\tau}.
\end{align}
\end{subequations}

These relations show that cross-polarized invisibility inherently requires bianisotropy, since purely electric and magnetic responses cannot simultaneously suppress reflection and co-polarized transmission under asymmetric and oblique-incidence conditions.
If the appropriate set of conditions is satisfied for a given incident polarization, the metasurface cancels reflection, suppresses co-polarized transmission, and converts the entire transmitted field into the orthogonal polarization without adding phase. The structure is therefore electromagnetically invisible while simultaneously acting as a lossless polarization converter.
These two sets of relations can, in principle, be satisfied by a single metasurface exhibiting a Lorentzian-type resonance, whose dispersive behavior allows the susceptibilities to take both positive and negative values over frequency. In such a scenario, the metasurface could fulfill the TE-to-TM polarization conversion condition at one frequency and the TM-to-TE conversion condition at a different frequency.

To demonstrate the validity of relations~\eqref{eq_TM_to_TE} and~\eqref{eq_TE_to_TM}, we respectively plot in Fig.~\ref{fig_Tcx}(c) and Fig.~\ref{fig_Tcx}(d) the cross-polarized transmittance and transmission phase, assuming a TM-polarized obliquely propagating incident wave, for varying values of $\chiem{xx}$ and $\chiem{yy}$. In Fig.~\ref{fig_Tcx}(c), we see two regions where $|T_\text{cx}|^2=1$, however, when looking at the phase, we see that only one of them corresponds to a zero-phase shift. This region is highlighted by a star that corresponds to the invisibility condition~\eqref{eq_TE_to_TM}.

\section{Full-Wave Validation}

We now simulate a metasurface embedded in dissimilar media to validate the conditions previously derived. We concentrate our attention on the case of co-polarized invisibility. The design of the simulated structure is kept consistent with current fabrication capabilities and the materials are chosen to be realistic, with material parameters obtained from ellipsometric measurements performed in-house. We design the structure to achieve an effective asymmetry along the normal direction. This asymmetry is not intrinsic to the structure itself but arises from the dissimilar surrounding media, which effectively breaks mirror symmetry along the normal direction, as discussed previously. The simulated unit-cell is an amorphous silicon cylinder lying on a $\text{SiO}_2$ substrate and with a vacuum superstrate. 

\begin{figure}[H]
    \centering
    \includegraphics[width=1\linewidth]{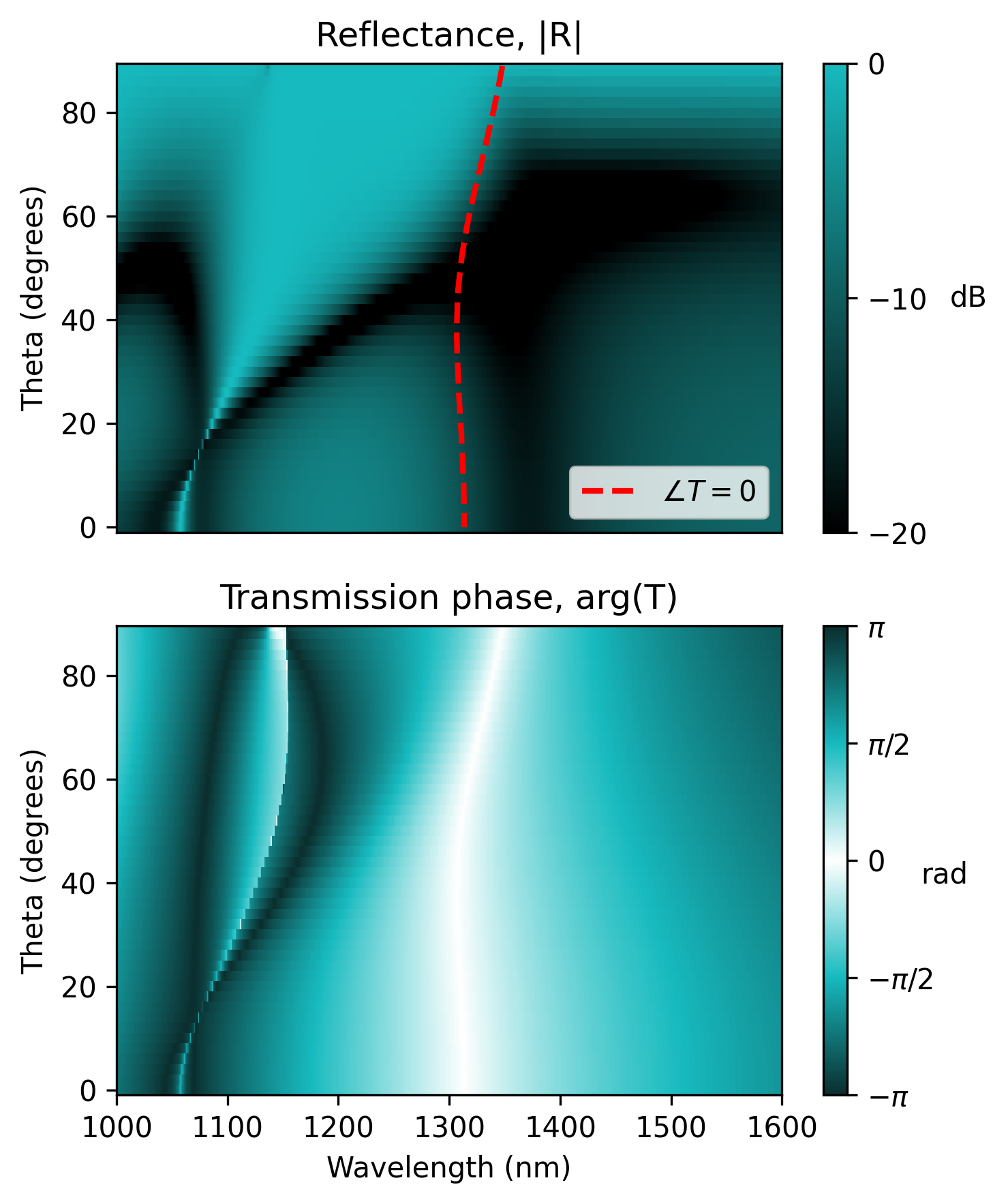}
    \caption{Full-wave simulated reflectance (top figure) and transmission phase (bottom figure) versus wavelength and incidence angle, $\theta$. The dashed line represents the zero transmission phase. The invisibility occurs at the intersection of the dashed line with the zero-reflectance region.}
    \label{simulation_co_invisibility}
\end{figure}

The metasurface is simulated with CST Studio and the resulting reflectance and transmission phase are plotted in Fig.~\ref{simulation_co_invisibility}. The dashed line in Fig.~\ref{simulation_co_invisibility} denotes the contour of the zero transmission phase. We clearly see that the zero transmission phase coincides with a zero of reflectance (less than $-20$~dB) around $45^\circ$ of incidence angle. Since the metasurface is essentially lossless in this wavelength range, we directly deduce that the zero of reflectance corresponds to full transmittance. In that region, we therefore achieve a relatively broadband invisibility.

\section{Conclusion}

We have shown that dipolar metasurfaces embedded within similar media cannot achieve non-trivial invisibility at normal incidence, but that this limitation is fundamentally lifted under oblique illumination through the coupling of tangential and normal components of electric and magnetic dipoles. Furthermore, we have also confirmed that invisibility at normal incidence is also achievable when including multipolar contributions, as was previously shown in the literature.

In asymmetric media, our results obtained through the GSTC for a dipolar metasurface, show that invisibility is governed by magnetoelectric coupling arising from interactions at oblique incidence. This coupling becomes essential for reciprocal, lossless and passive systems, implying that purely electric and magnetic responses are insufficient and that bianisotropy is fundamentally required. Specifically, we discuss two special cases of invisibility that depend only on the tangential components of purely bianisotropic susceptibilities leading to both co- and cross-polarization invisibility.

While the found conditions of invisibility requiring purely bianisotropic responses might a priori seem impossible to practically implement, we conveniently showed that such kind of bianisotropy can emerge in an anisotropic metasurface from environmental asymmetry. This allows obtaining effective bianisotropic metasurface emerging from the media asymmetry avoiding complex unit-cell nanofabrication. 

We have further validated our results with full-wave simulations of nanofabrication friendly metasurfaces that show that invisibility can be achieved in realistic metasurface configurations without requiring intrinsically bianisotropic unit-cells. Experimental validation of the proposed framework is currently in progress and will be reported in future work.

\begin{acknowledgments}
We acknowledge funding from the Swiss National Science Foundation (project TMSGI2\_218392).
\end{acknowledgments}

\bibliography{apssamp}

\clearpage
\onecolumngrid
 
\appendix
\renewcommand{\thefigure}{\thesection.\arabic{figure}}
\setcounter{figure}{0}
\section*{Supplementary Information}

\section{Unitary Transmission Representation}

In this section, we show that the transmission through a lossless and reflectionless metasurface can be expressed as a unitary transformation acting on the tangential electric field, once the power normalization imposed by the surrounding media is accounted for.

We consider a metasurface separating two homogeneous media characterized by wave impedances $\eta_1$ and $\eta_2$, and supporting an incident plane wave at an angle $\theta_1$, transmitted at an angle $\theta_2$. Under the invisibility condition, the reflected field vanishes and the transmitted field may be written as
\begin{equation}
\mathbf{E}_\mathrm{t} = \overline{\overline{T}} \cdot \mathbf{E}_\mathrm{i},
\end{equation}
where $\overline{\overline{T}}$ is the transmission dyadic.

\subsection{Power normalization}

For a lossless metasurface, the normal component of the time-averaged Poynting vector must be conserved across the interface,
\begin{equation}
\langle S_{z,1} \rangle = \langle S_{z,2} \rangle.
\end{equation}
For a plane wave in medium $m$, the normal power flux is given by
\begin{equation}
\langle S_{z,m} \rangle = \frac{|\mathbf{E}_m|^2}{2\eta_m}\cos\theta_m,
\end{equation}
which leads to the relation
\begin{equation}
\frac{|\mathbf{E}_\mathrm{i}|^2}{\eta_1}\cos\theta_1=\frac{|\mathbf{E}_\mathrm{t}|^2}{\eta_2}\cos\theta_2.
\end{equation}
This condition imposes that the transmission dyadic satisfies
\begin{equation}
\label{eq_TTI}
\overline{\overline{T}}^\dagger
\left(\frac{\cos\theta_2}{\eta_2}\right)
\overline{\overline{T}}=\left(\frac{\cos\theta_1}{\eta_1}\right)
\overline{\overline{I}}.
\end{equation}

\subsection{Unitary form of the transmission}

It is convenient to factor out the scalar normalization term associated with the surrounding media in~\eqref{eq_TTI}. Defining
\begin{equation}
\tau = \sqrt{\frac{\eta_2 \cos\theta_1}{\eta_1 \cos\theta_2}},
\end{equation}
we write the transmission dyadic as
\begin{equation}
\overline{\overline{T}} = \tau \overline{\overline{U}}.
\end{equation}
Substituting into the previous relation yields
\begin{equation}
\overline{\overline{U}}^\dagger \overline{\overline{U}} = \overline{\overline{I}},
\end{equation}
which shows that $\overline{\overline{U}}$ is a unitary matrix.

Therefore, once the power normalization imposed by the surrounding media is taken into account, the metasurface acts as a unitary transformation on the polarization state of the incident field.

\subsection{Physical interpretation}

The unitary matrix $\overline{\overline{U}}$ describes all possible lossless polarization transformations supported by the metasurface. In particular,
\begin{itemize}
\item $\overline{\overline{U}} = \overline{\overline{I}}$ corresponds to co-polarized transmission without modification of the polarization state,
\item off-diagonal unitary matrices correspond to complete cross-polarization conversion,
\item more general unitary matrices describe rotations or mixing of polarization components.
\end{itemize}
This formulation provides a unified description of co- and cross-polarized invisible responses, which are treated separately in the main text. It also clarifies that, under lossless and reflectionless conditions, the metasurface cannot introduce any net gain or loss in the polarization space, and is therefore restricted to unitary transformations.

\section{Derivation of the Generalized GSTC}
\subsection{Dipole GSTC}
These relations describe the discontinuity of the tangential electromagnetic fields across an infinitesimally thin metasurface in terms of the induced electric and magnetic surface current densities. The gradient terms account for the contribution of normal polarization components, which become particularly important under oblique incidence where tangential phase variations are present.
\begin{subequations} \label{gstc_supp}
\begin{align}
    \DvHp &= - j\omega\zhat \times \Pp - \frac{1}{ \mu_0} \nabla_{\parallel} M_z, \label{gstc_a_supp}\\
    \DvEp &= + j\omega\zhat \times \Mp - \frac{1}{ \epsilon_0} \nabla_{\parallel} P_z \label{gstc_b_supp},
\end{align}
\end{subequations}
where $\mathbf{P}$ and $\mathbf{M}$ are the electric and magnetic polarizations densities respectively, and $\DvHp$ and $\DvEp$ are the tangential magnetic and electric field differences.
\subsubsection*{Tangential electric and magnetic fields}
In a general case, the field differences are defined as
\begin{subequations} \label{eq_deltas}
\begin{align}
\DvHp &= \mathbf{H}_{\parallel,\text{t}} - \mathbf{H}_{\parallel,\text{r}} - \mathbf{H}_{\parallel,\text{i}},\\
\DvEp &= \mathbf{E}_{\parallel,\text{t}} - \mathbf{E}_{\parallel,\text{r}} - \mathbf{E}_{\parallel,\text{i}},
\end{align}
\end{subequations}
where the subscripts \{i,t,r\} denote the incident, transmitted and reflected fields. Note that invisibility always requires that $\mathbf{E}_\text{r} = \mathbf{H}_\text{r} = 0$.

We consider no reflection field, thus the electric and magnetic jumps for the tangential and normal components can be expressed as,
\begin{equation}
\begin{aligned}
\Delta \vEp 
&= \vEt - \vEi \\[4pt]
&= \ktwo\,\Tmat \cdot \vEi - \kone\,\Id \cdot \vEi \\[4pt]
&= \left( \ktwo\,\Tmat - \kone\,\Id \right) \cdot \vEi\\[4pt]
&= \DeltaT \cdot \vEi.
\end{aligned}
\end{equation}

\begin{equation}
\begin{aligned}
\DeltaH
&= \vHt - \vHi \\[4pt]
&= \pm\,\zhat \times 
\left[
\Ytwo \cdot \vEt
- \Yone \cdot \kone\,\vEi
\right] \\[4pt]
&= \pm\,\zhat \times 
\left[
\left(
\ktwo\,\Ytwo \cdot \Tmat
- \kone\,\Yone
\right)
\cdot \vEi
\right]\\[6pt]
&=\pm\,\zhat \times
\left [
\Ydel \cdot \vEi
\right ]\\[6pt]
&=\pm \,\Rot\cdot\Ydel \cdot \vEi.
\end{aligned}
\end{equation}
These expressions relate the field discontinuities directly to the transmission dyadic of the metasurface under the assumption of zero reflection. The operators $\DeltaT$ and $\Ydel$ therefore play the role of effective transfer and admittance tensors that compactly describe how the metasurface modifies the incident field.
We define the $\kappa_i$ coefficients as,
\begin{equation}
    \begin{cases}
    \kappa_1 = \cos\theta_1 \quad\text{and}\quad  \kappa_2 = \cos\theta_2 \quad\text{for TM waves,}\\
    \kappa_1 = \kappa_2 = 1 \quad\text{for TE waves}.
    \end{cases}
\end{equation}
These factors arise from the projection of the wavevector onto the surface normal and account for the polarization-dependent relationship between tangential and normal field components. Their inclusion ensures that both TE and TM polarizations are treated consistently within the same formalism.
The upper sign of the definitions corresponds to forward incidence from medium 1 to medium 2, and the lower sign corresponds to reverse incidence. \newline
\subsubsection*{Admittance tensor definition}
The general admittance tensor is expressed as it follows~\cite{tretyakov2003},
\begin{align}
\overline{\overline{Y}} &=
\underbrace{\frac{\omega \varepsilon}{k_z} \,
\frac{\kpar \kpar}{\kpar^2}}_{\text{TM}}
\;+\;
\underbrace{\frac{k_z}{\omega \mu} \,
\frac{\hat{\mathbf{z}} \times \kpar \; \hat{\mathbf{z}} \times \kpar}{\kpar^2}}_{\text{TE}}.
\end{align}
The admittance tensor provides a compact way of expressing the relationship between the tangential electric and magnetic fields of a plane wave. Its decomposition into TM and TE contributions highlights the different physical origins of the field components associated with longitudinal and transverse wave impedance. \newline
\subsubsection*{Surface polarization densities and average fields}
The electric and magnetic surface polarization densities are related to the averaged electromagnetic fields at the metasurface through the susceptibility tensors. These tensors include the bianisotropic response of the structure and include both tangential and normal coupling mechanisms.
\begin{align}
\Pp &= \varepsilon_0 \left( \Xeepp
 \cdot \vEpav  
+ \Xeepz \Ezav \right)
+ \frac{1}{c_0} \left( 
\Xempp \cdot \vHpav
+ \Xempz \Hzav \right), \\[1em]
\Mp &= \mu_0 \left( 
\Xmmpp \cdot \vHpav 
+ \Xmmpz \Hzav \right)
+ \frac{1}{c_0} \left( 
\Xmepp \cdot \vEpav
+ \Xmepz \Ezav \right), \\[1em]
P_z &= \epsz \left( 
\Xeezp \cdot \vEpav
+ \Xeezz \Ezav \right)
+ \frac{1}{c_0} \left( 
\Xemzp \cdot \vHpav
+ \Xemzz \Hzav \right), \\[1em]
M_z &= \mu_0 \left( 
\Xmmzp \cdot \vHpav 
+ \Xmmzz \Hzav \right)
+ \frac{1}{c_0} \left(
\Xmezp \cdot \vEpav
+ \Xmezz \Ezav \right).
\end{align}

The electric and magnetic polarization densities may be decomposed into tangential ($\parallel$) and normal ($z$) components and related to the averaged tangential and normal electric and magnetic fields through surface-susceptibility tensors as
\begin{equation}
\label{pol_matrix}
\begin{bmatrix}
\Pp\\[2pt]
\Pz\\[6pt]
\Mp\\[2pt]
\Mz
\end{bmatrix}
=
\begin{bmatrix}
\epsz\,\Xeepp & \epsz\,\Xeepz & \dfrac{1}{c_0}\,\Xempp & \dfrac{1}{c_0}\,\Xempz\\[8pt]
\epsz\,\Xeezp & \epsz\,\Xeezz & \dfrac{1}{c_0}\,\Xemzp & \dfrac{1}{c_0}\,\Xemzz\\[10pt]
\dfrac{1}{c_0}\,\Xmepp & \dfrac{1}{c_0}\,\Xmepz & \mu_0\,\Xmmpp & \mu_0\,\Xmmpz\\[8pt]
\dfrac{1}{c_0}\,\Xmezp & \dfrac{1}{c_0}\,\Xmezz & \mu_0\,\Xmmzp & \mu_0\,\Xmmzz
\end{bmatrix}
\begin{bmatrix}
\vEpav\\[2pt]
\Ezav\\[6pt]
\vHpav\\[2pt]
\Hzav
\end{bmatrix},
\end{equation}
where $\mathbf{E}_\text{av}$ and $\mathbf{H}_\text{av}$ are the electric and magnetic field averages defined as
\begin{subequations} \label{eq_averages}
\begin{align}
\mathbf{E}_\text{av} &= \frac{1}{2}\left(\mathbf{E}_{\text{i}} + \mathbf{E}_{\text{r}} + \mathbf{E}_{\text{t}}\right),\\
\mathbf{H}_\text{av} &= \frac{1}{2}\left(\mathbf{H}_{\text{i}} + \mathbf{H}_{\text{r}} + \mathbf{H}_{\text{t}}\right).
\end{align}
\end{subequations}

In \eqref{pol_matrix}, the size of each susceptibility block is determined by its upper indices:
a block with upper index $\parallel\parallel$ is a $2\times2$ dyadic,
a block with upper index $\parallel z$ is a $2\times1$ column block,
a block with upper index $z\parallel$ is a $1\times2$ row block,
and a block with upper index $zz$ is a scalar.

The susceptibility tensors are defined as follows:
\begin{align}
\Xee &=
\left(
\begin{array}{c|c}
\Xeepp & \Xeepz \\ \hline
\Xeezp & \Xeezz
\end{array}
\right)
=
\left(
\begin{array}{cc|c}
\chiee{xx} & \chiee{xy} & \chiee{xz} \\
\chiee{yx} & \chiee{yy} & \chiee{yz} \\ \hline
\chiee{zx} & \chiee{zy} & \chiee{zz}
\end{array}
\right), \\[1em]
\Xem &=
\left(
\begin{array}{c|c}
\Xempp & \Xempz \\ \hline
\Xemzp & \Xemzz
\end{array}
\right)
=
\left(
\begin{array}{cc|c}
\chiem{xx} & \chiem{xy} & \chiem{xz} \\
\chiem{yx} & \chiem{yy} & \chiem{yz} \\ \hline
\chiem{zx} & \chiem{zy} & \chiem{zz}
\end{array}
\right), \\[1em]
\Xme &=
\left(
\begin{array}{c|c}
\Xmepp & \Xmepz \\ \hline
\Xmezp & \Xmezz
\end{array}
\right)
=
\left(
\begin{array}{cc|c}
\chime{xx} & \chime{xy} & \chime{xz} \\
\chime{yx} & \chime{yy} & \chime{yz} \\ \hline
\chime{zx} & \chime{zy} & \chime{zz}
\end{array}
\right), \\[1em]
\Xmm &=
\left(
\begin{array}{c|c}
\Xmmpp & \Xmmpz \\ \hline
\Xmmzp & \Xmmzz
\end{array}
\right)
=
\left(
\begin{array}{cc|c}
\chimm{xx} & \chimm{xy} & \chimm{xz} \\
\chimm{yx} & \chimm{yy} & \chimm{yz} \\ \hline
\chimm{zx} & \chimm{zy} & \chimm{zz}
\end{array}
\right).
\end{align}

Expressing the average fields in terms of the admittance tensors leads to,
\begin{equation}
\begin{aligned}
\vEpav
&= \frac{1}{2}(\vEt + \vEi) \\[4pt]
&= \frac{1}{2}\left ( \ktwo\,\Tmat \cdot \vEi + \kone\,\Id \cdot \vEi \right ) \\[4pt]
&= \left( \ktwo\,\Tmat + \kone\,\Id \right) \cdot \frac{\vEi}{2}\\[4pt]
&= \AlphaT \cdot \frac{\vEi}{2}.
\end{aligned}
\end{equation}

% === Tangential Magnetic Field Average ===
\begin{equation}
\begin{aligned}
\vHpav
&= \frac{1}{2}\big(\vHt + \vHi\big) \\[4pt]
&= \pm\,\frac{1}{2}\,\zhat \times 
\left[
\Ytwo \cdot \vEt
+ \Yone \cdot \kone\,\vEi
\right] \\[4pt]
&= \pm\,\frac{1}{2}\,\zhat \times 
\left[
\left(
\ktwo\,\Ytwo \cdot \Tmat
+ \kone\,\Yone
\right)\cdot \vEi
\right] \\[6pt]
&= \pm\,\frac{1}{2}\,\zhat \times\big(\Ysig \cdot \vEi\big) \\[6pt]
&= \pm\,\Rot\cdot\Ysig \cdot \frac{\vEi}{2}.
\end{aligned}
\end{equation}

% === Normal Electric Field Average ===
\begin{equation}
\begin{aligned}
\Ezav 
&= \frac{1}{2}\big(\ertwo E_{z2} + \erone E_{z1}\big) \\[4pt]
&= \mp\,\frac{1}{2}\,\frac{1}{\omega \epsz}
\left[
\kpar \cdot \Ytwo \cdot \vEt
+
\kpar \cdot \Yone \cdot \kone\,\vEi
\right] \\[6pt]
&= \mp\,\frac{1}{2}\,\frac{\kpar}{\omega \epsz} \cdot
\left[
\ktwo\,\Ytwo \cdot \Tmat
+
\kone\,\Yone
\right]\cdot \vEi \\[6pt]
&= \mp\,\frac{\kpar}{\omega \epsz} \cdot \Ysig \cdot \frac{\vEi}{2}.
\end{aligned}
\end{equation}

% === Normal Magnetic Field Average ===
\begin{equation}
\begin{aligned}
\Hzav 
&= \frac{1}{2}(\murtwo H_{z2} + \murone H_{z1}) \\[6pt]
&= \frac{1}{2}\frac{1}{\omega \mu_0}
\left\{
\,\zhat \cdot \big[ \kpar \times (\ktwo\,\Tmat \cdot \vEi) \big]
\;+\;\zhat \cdot \big[ \kpar \times (\kone\,\vEi) \big]
\right\} \\[8pt]
&=\frac{1}{2\omega \mu_0} \left\{ (\zhat \times \kpar ) \cdot \left[ (\ktwo\,\Tmat \cdot \vEi) + (\kone\, \cdot \vEi) \right] \right\} \\[8pt]
&=\frac{1}{2\omega \mu_0} \left\{ (\zhat \times \kpar ) \cdot \left[ \ktwo\,\Tmat + \kone\, \Id \right] \right\}\cdot \vEi \\[8pt]
&= \frac{1}{\omega\mu_0}\; \Rot \cdot \kpar \cdot \AlphaT \cdot \frac{\vEi}{2}.
\end{aligned}
\end{equation}
Where $z$-rotation matrix (i.e. $\mathbf{\hat{z}}\times$) is defined as,
\begin{align}
\Rot =
\begin{bmatrix}
    0 & -1 \\
    1 & 0
\end{bmatrix}.
\end{align}

The average field are only in terms of the incident field and the transmission dyadic. As a result, the surface polarizations can be written as explicit functions of the incident field, which is a key step toward deriving closed-form invisibility conditions. \newline
For compactness, we introduce the following effective tensors,
\begin{align}
\label{eq_dyadic_def}
\AlphaT &\;\equiv\; \kone\,\Id \;+\; \ktwo\,\Tmat, \\[4pt]
\Ysig   &\;\equiv\; \kone\,\Yone \;+\; \ktwo\,\Ytwo\cdot\Tmat, \\[4pt]
\Ydel   &\;\equiv\; \ktwo\,\Ytwo\cdot\Tmat \;-\; \kone\,\Yone, 
% \Ysig   &\;\equiv\; \kone\,\Yone \;+\; \ktwo\,\Ytwo\cdot\Tmat, \\[4pt]
% \AlphaT    &\;\equiv\; \ktwo\,\Tmat
% + \kone\,\Id.
\end{align}
which naturally emerge in the GSTC formulation when the averaged fields are substituted into the polarization relations. These quantities encapsulate the influence of the surrounding media and the transmission response of the metasurface.
Substituting the average fields into the electric and magnetic surface polarization densities,
\begin{equation}
\begin{aligned}
\Pp
&=
\varepsilon_0 \Bigg\{
\Xeepp \cdot \left( \AlphaT \cdot \frac{\vEi}{2} \right)
+ \Xeepz
\left[ \mp \frac{\kpar}{\omega \epsz} \cdot \Ysig \cdot \frac{\vEi}{2} \right]
\Bigg\} \\
&\quad + \frac{1}{c_0} \Bigg\{
\Xempp \cdot
\left[ \pm \Rot \cdot \Ysig \cdot \frac{\vEi}{2} \right]
+ \Xempz
\left[ \frac{1}{\omega \mu_0} \Rot \cdot \kpar \cdot \AlphaT \cdot \frac{\vEi}{2} \right]
\Bigg\}, \\[1em]
\end{aligned}
\end{equation}
\begin{equation}
\begin{aligned}
\Mp
&=
\mu_0 \Bigg\{
\Xmmpp \cdot
\left[ \pm \Rot \cdot \Ysig \cdot \frac{\vEi}{2} \right]
+ \Xmmpz
\left[ \frac{1}{\omega \mu_0} \Rot \cdot \kpar \cdot \AlphaT \cdot \frac{\vEi}{2} \right]
\Bigg\} \\
&\quad + \frac{1}{c_0} \Bigg\{
\Xmepp \cdot \left( \AlphaT \cdot \frac{\vEi}{2} \right)
+ \Xmepz
\left[ \mp \frac{\kpar}{\omega \epsz} \cdot \Ysig \cdot \frac{\vEi}{2} \right]
\Bigg\}, \\[1em]
\end{aligned}
\end{equation}
\begin{equation}
\begin{aligned}
\Pz
&=
\varepsilon_0 \Bigg\{
\Xeezp \cdot \left( \AlphaT \cdot \frac{\vEi}{2} \right)
+ \Xeezz
\left[ \mp \frac{\kpar}{\omega \epsz} \cdot \Ysig \cdot \frac{\vEi}{2} \right]
\Bigg\} \\
&\quad + \frac{1}{c_0} \Bigg\{
\Xemzp \cdot
\left[ \pm \Rot \cdot \Ysig \cdot \frac{\vEi}{2} \right]
+ \Xemzz
\left[ \frac{1}{\omega \mu_0} \Rot \cdot \kpar \cdot \AlphaT \cdot \frac{\vEi}{2} \right]
\Bigg\}, \\[1em]
\end{aligned}
\end{equation}
\begin{equation}
\begin{aligned}
\Mz
&=
\mu_0 \Bigg\{
\Xmmzp \cdot
\left[ \pm \Rot \cdot \Ysig \cdot \frac{\vEi}{2} \right]
+ \Xmmzz
\left[ \frac{1}{\omega \mu_0} \Rot \cdot \kpar \cdot \AlphaT \cdot \frac{\vEi}{2} \right]
\Bigg\} \\
&\quad + \frac{1}{c_0} \Bigg\{
\Xmezp \cdot \left( \AlphaT \cdot \frac{\vEi}{2} \right)
+ \Xmezz
\left[ \mp \frac{\kpar}{\omega \epsz} \cdot \Ysig \cdot \frac{\vEi}{2} \right]
\Bigg\}.
\end{aligned}
\end{equation}
\subsubsection*{Effective susceptibilities}
By substituting the average fields into the polarization--magnetization relations, one obtains \emph{effective} susceptibilities 
$\chi^{\mathrm{eff}}$ that provide a direct algebraic link between the desired scattering properties and 
the required metasurface parameters.
% === Tangential Electric Polarization ===
\begin{equation}
\begin{aligned}
\mathbf{P}_\parallel 
&= 
\left( 
\epsz \, \Xeepeff 
+ \frac{1}{c_0} \, \Xempeff 
\right) \cdot \vEi, 
\quad\text{where}\quad
\left\{
\begin{aligned}
\Xeepeff
&\equiv 
\frac{1}{2}\Bigg\{
\Xeepp \cdot \AlphaT 
\mp 
\Xeepz \otimes
\!\left[
\frac{1}{\omega \epsz}\,
\kpar \cdot \Ysig
\right]\!\Bigg\}, \\[0.6em]
\Xempeff
&\equiv 
\frac{1}{2}\Bigg\{
\pm\,\Xempp \cdot 
\left[\Rot \cdot \Ysig\right]
+ 
\Xempz \otimes
\!\left[
\frac{1}{\omega\mu_0}\; \Rot \cdot \kpar \cdot \AlphaT
\right]\!\Bigg\},
\end{aligned}
\right.
\end{aligned}
\end{equation}

% === Tangential Magnetic Polarization ===
\begin{equation}
\begin{aligned}
\mathbf{M}_\parallel 
&= 
\left( 
\mu_0 \, \Xmmpeff 
+ \frac{1}{c_0} \, \Xmepeff 
\right) \cdot \vEi, 
\quad\text{where}\quad
\left\{
\begin{aligned}
\Xmmpeff
&\equiv 
\frac{1}{2}\Bigg\{
\pm\,\Xmmpp \cdot 
\left[\Rot \cdot \Ysig\right]
+ 
\Xmmpz \otimes
\!\left[
\frac{1}{\omega\mu_0}\; \Rot \cdot \kpar \cdot \AlphaT
\right]\!\Bigg\}, \\[0.6em]
\Xmepeff
&\equiv 
\frac{1}{2}\Bigg\{
\Xmepp \cdot \AlphaT
\mp 
\Xmepz \otimes
\!\left[
\frac{1}{\omega \epsz}\,
\kpar \cdot \Ysig
\right]\!\Bigg\},
\end{aligned}
\right.
\end{aligned}
\end{equation}

%
% === Normal Electric Polarization ===
\begin{equation}
\begin{aligned}
P_z 
&= 
\left( 
\epsz \, \Xeezeff 
+ \frac{1}{c_0} \, \Xemzeff 
\right) \cdot \vEi,
\quad\text{where}\quad
\left\{
\begin{aligned}
\Xeezeff
&\equiv 
\frac{1}{2}\Bigg\{
\Xeezp \cdot \AlphaT 
\mp 
\Xeezz
\!\left[
\frac{1}{\omega \epsz}\,
\kpar \cdot \Ysig
\right]\!\Bigg\}, \\[0.6em]
\Xemzeff
&\equiv 
\frac{1}{2}\Bigg\{
\pm\,\Xemzp \cdot 
\left[\Rot \cdot \Ysig\right]
+
\Xemzz
\!\left[
\frac{1}{\omega\mu_0}\; \Rot \cdot \kpar \cdot \AlphaT
\right]\!\Bigg\},
\end{aligned}
\right.
\end{aligned}
\end{equation}

%
% === Normal Magnetic Polarization ===
\begin{equation}
\begin{aligned}
M_z 
&= 
\left( 
\mu_0 \, \Xmmzeff 
+ \frac{1}{c_0} \, \Xmezeff 
\right) \cdot \vEi,
\quad\text{where}\quad
\left\{
\begin{aligned}
\Xmmzeff
&\equiv 
\frac{1}{2}\Bigg\{
\pm\,\Xmmzp \cdot 
\left[\Rot \cdot \Ysig\right]
+
\Xmmzz
\!\left[
\frac{1}{\omega\mu_0}\; \Rot \cdot \kpar \cdot \AlphaT
\right]\!\Bigg\}, \\[0.6em]
\Xmezeff
&\equiv 
\frac{1}{2}\Bigg\{
\Xmezp \cdot \AlphaT
\mp 
\Xmezz
\!\left[
\frac{1}{\omega \epsz}\,
\kpar \cdot \Ysig
\right]\!\Bigg\},
\end{aligned}
\right.
\end{aligned}
\end{equation}
where the effective susceptibilities are rewritten as,
\begin{subequations}\label{chieff}
\begin{equation}
\Xiepeff
= \frac{1}{2}\Big\{
\Xiepp\cdot \AlphaT
\mp \Xiepz\otimes
\Big[\frac{1}{\omega\varepsilon_0}\,\kpar\cdot \Ysig\Big]
\Big\},
\end{equation}

\begin{equation}
\begin{aligned}
\Ximpeff
= \frac{1}{2}\Big\{&
\pm \Ximpp \cdot (\Rot\cdot \Ysig)
+ \Ximpz \otimes
\Big[\frac{1}{\omega\mu_0}\,\Rot\cdot \kpar \cdot \AlphaT\Big]
\Big\},
\end{aligned}
\end{equation}

\begin{equation}
\Xiezeff
= \frac{1}{2}\Big\{
\bm{\chi}_\mathrm{ie}^{z\parallel}\cdot \AlphaT
+ \chi_\mathrm{ie}^{zz}
\Big[\frac{1}{\omega\varepsilon_0}\,\kpar\cdot \Ysig\Big]
\Big\},
\end{equation}

\begin{equation}
\begin{aligned}
\Ximzeff
= \frac{1}{2}\Big\{&
\pm \bm{\chi}_\mathrm{im}^{z\parallel}\cdot (\Rot\cdot \Ysig)
+ \chi_\mathrm{im}^{zz}
\Big[\frac{1}{\omega\mu_0}\,\Rot\cdot \kpar \cdot \AlphaT\Big]
\Big\},
\end{aligned}
\end{equation}
\end{subequations}
with $i \in \{\mathrm{e,m}\}$.

To summarize, under reflection cancellation ($R=0$), the field differences expressed in terms of the incident electric field read 
\begin{align}
\label{field_jump}
\DvEp &= \DeltaT \cdot \vEi, \\
\DvHp &= \pm \Rot \cdot \Ydel \cdot \vEi,
\end{align}
where $\DeltaT$, $\Ydel$ and $\Rot$ are the differentiated transfer dyadic, substracted admittance tensor and the $z$-oriented rotation matrix. Similarly, the field averages are expressed as
\begin{subequations}\label{avg_fields}
\begin{align}
    \vEpav &= \AlphaT \cdot \frac{\vEi}{2}, \label{avg_fields_a}\\
    \vHpav &= \pm\, \Rot \cdot \Ysig \cdot \frac{\vEi}{2}, \label{avg_fields_b}\\
    \Ezav  &= \mp\, \frac{\kpar}{\omega \epsz} \cdot \Ysig \cdot \frac{\vEi}{2}, \label{avg_fields_c}\\
    \Hzav  &= \frac{1}{\omega\mu_0}\, \Rot \cdot \kpar \cdot \AlphaT \cdot \frac{\vEi}{2}, \label{avg_fields_d}
\end{align}
\end{subequations}
where $\AlphaT$ is the effective transfer dyadic, $\Ysig$ is sum admittance tensors.

Substituting the field averages~\eqref{avg_fields} into the constitutive relation~\eqref{pol_matrix} leads to
\begin{subequations}\label{eff_pol}
\begin{align}
    \Pp &= \left( \epsz\Xeepeff + \frac{1}{c_0}\Xempeff \right)\cdot \vEi, \label{eff_pol_a}\\
    \Mp &= \left( \mu_0\Xmmpeff + \frac{1}{c_0}\Xmepeff \right)\cdot \vEi, \label{eff_pol_b}\\
    \Pz &= \left( \epsz\Xeezeff + \frac{1}{c_0}\Xemzeff \right)\cdot \vEi, \label{eff_pol_c}\\
    \Mz &= \left( \mu_0\Xmmzeff + \frac{1}{c_0}\Xmezeff \right)\cdot \vEi. \label{eff_pol_d}
\end{align}
\end{subequations}

\subsubsection*{Reciprocity}
The reciprocity condition imposes that the electric and magnetic susceptibility tensors are symmetric, while the magnetoelectric coupling tensors are related through an antisymmetric transpose relation as
\begin{align}
\label{eq_recip}
\overline{\overline{\chi}}_\mathrm{ee}=\overline{\overline{\chi}}_\mathrm{ee}^T\,, \quad 
\overline{\overline{\chi}}_\mathrm{mm}=\overline{\overline{\chi}}_\mathrm{mm}^T\,, \quad
\overline{\overline{\chi}}_\mathrm{em}=-\overline{\overline{\chi}}_\mathrm{me}^T.
\end{align}

\subsubsection*{Dipole GSTC for dissimilar media}
Inserting the compact form of the electric and magnetic surface polarization densities in~\eqref{eff_pol} into the GSTC equations, we obtain
\begin{equation}
\begin{aligned}
\pm\,\Rot\cdot\Ydel\cdot \vEi
&= -\,j\omega\,\Rot\cdot\Bigg\{
\epsz\,\frac{1}{2}\bigg[
\Xeepp\!\cdot\!\AlphaT
\mp \Xeepz\;\otimes
\frac{1}{\omega\epsz}\,
\mathbf{k}_\parallel\!\cdot\!\Ysig
\bigg] \\
&\qquad\qquad\qquad\qquad
+ \frac{1}{c_0}\,\frac{1}{2}\bigg[
\pm\,\Xempp\!\cdot\!(\Rot\!\cdot\!\Ysig)
+ \Xempz\;\otimes
\frac{1}{\omega\mu_0}\,(\Rot\!\cdot\!\mathbf{k}_\parallel)\cdot\AlphaT
\bigg]
\Bigg\}\cdot \vEi\\[6pt]
&\quad - \frac{1}{\mu_0}\,\boldsymbol{\nabla}_\parallel
\Bigg\{
\mu_0\,\frac{1}{2}\bigg[
\pm\,\Xmmzp\!\cdot\!(\Rot\!\cdot\!\Ysig)
+ \Xmmzz\;
\frac{1}{\omega\mu_0}\,(\Rot\!\cdot\!\mathbf{k}_\parallel)\cdot\AlphaT
\bigg] \\
&\qquad\qquad\qquad\qquad
+ \frac{1}{c_0}\,\frac{1}{2}\bigg[
\Xmezp\!\cdot\!\AlphaT
\mp \Xmezz\;
\frac{1}{\omega\epsz}\,
\mathbf{k}_\parallel\!\cdot\!\Ysig
\bigg]
\Bigg\}\cdot \vEi.
\end{aligned}
\end{equation}

\begin{equation}
\begin{aligned}
\DeltaT\,\cdot \vEi
&= +\,j\omega\,\Rot\cdot\Bigg\{
\mu_0\,\frac{1}{2}\bigg[
\pm\,\Xmmpp\!\cdot\!(\Rot\!\cdot\!\Ysig)
+ \Xmmpz\;\otimes
\frac{1}{\omega\mu_0}\,(\Rot\!\cdot\!\mathbf{k}_\parallel)\cdot\AlphaT
\bigg] \\
&\qquad\qquad\qquad\qquad
+ \frac{1}{c_0}\,\frac{1}{2}\bigg[
\Xmepp\!\cdot\!\AlphaT
\mp \Xmepz\;\otimes
\frac{1}{\omega\epsz}\,
\mathbf{k}_\parallel\!\cdot\!\Ysig
\bigg]
\Bigg\}\cdot \vEi\\[6pt]
&\quad - \frac{1}{\epsz}\,\frac{1}{2}\,\boldsymbol{\nabla}_\parallel
\Bigg\{
\epsz\bigg[
\Xeezp\!\cdot\!\AlphaT
\mp \Xeezz\;
\frac{1}{\omega\epsz}\,
\mathbf{k}_\parallel\!\cdot\!\Ysig
\bigg] \\
&\qquad\qquad\qquad\qquad
+ \frac{1}{c_0}\,\frac{1}{2}\bigg[
\pm\,\Xemzp\!\cdot\!(\Rot\!\cdot\!\Ysig)
+ \Xemzz\;
\frac{1}{\omega\mu_0}\,(\Rot\!\cdot\!\mathbf{k}_\parallel)\cdot\AlphaT
\bigg]
\Bigg\}\cdot \vEi.
\end{aligned}
\end{equation}
We can express the expanded equation as a function of the effective susceptibilities to end up with the following compact form,
\begin{align}
\pm \,\Rot\cdot\Ydel \cdot \vEi
&= 
\Big[
- j \omega\, \Rot \cdot 
\big( \epsz\, \Xeepeff 
+ \tfrac{1}{c_0}\, \Xempeff \big)
\;-\;
\tfrac{1}{\mu_0}\, \boldsymbol{\nabla}_{\parallel} 
\big( \mu_0\, \Xmmzeff + \tfrac{1}{c_0}\, \Xmezeff \big)
\Big] \cdot \vEi,
\\[0.6em]
\DeltaT \cdot \vEi
&= 
\Big[
+ j \omega\, \Rot \cdot 
\big( \mu_0\, \Xmmpeff 
+ \tfrac{1}{c_0}\, \Xmepeff \big)
\;-\;
\tfrac{1}{\epsz}\, \boldsymbol{\nabla}_{\parallel}
\big( \epsz\, \Xeezeff + \tfrac{1}{c_0}\, \Xemzeff \big)
\Big] \cdot \vEi.
\end{align}
Imposing reciprocity and conservation of power, these equations simplify significantly and reveal that only the magnetoelectric coupling terms remain. The resulting invisibility conditions reduce to

\begin{subequations}
\label{eq_general_lossless}
\begin{equation}
\begin{aligned}
\pm\,2j\,\Rot\!\cdot\!\Ydel \cdot \vEi
&= \omega\,\Rot\!\cdot\!\Bigg\{
\frac{1}{c_0}\Big[
\pm\,\Xempp\!\cdot\!\big(\Rot\!\cdot\!\Ysig\big)
+ \Xempz\!\otimes\!
\frac{1}{\omega\mu_0}\,\AlphaT\cdot\big(\Rot\!\cdot\!\kpar\big)
\Big]
\Bigg\}  \cdot \vEi\\[4pt]
&\quad - \frac{1}{\mu_0}\,\kpar\!\otimes\!\Bigg\{
\frac{1}{c_0}\Big[
\big(-\,\Xempz\big)^{T}\!\cdot\!\AlphaT
\mp\,\big(-\Xemzz\big)\;
\frac{1}{\omega\epsz}\,\kpar\!\cdot\!\Ysig
\Big]
\Bigg\} \cdot \vEi.
\end{aligned}
\end{equation}
\begin{equation}
\begin{aligned}
    2 j\,\DeltaT \cdot \vEi &= -\omega \,\Rot\cdot \Bigg\{ \frac{1}{c_0}\Big[ \big(-\Xempp\big)^T \cdot \AlphaT \mp \big(-\Xempz \big)^T \otimes\frac{1}{\omega\epsz} \kpar \cdot \Ysig \Big] \Bigg\} \cdot \vEi \\[4pt]
    &\quad -\frac{1}{\epsz}\,\kpar\,\otimes\Bigg\{ \frac{1}{c_0} \Big [ \pm \Xemzp \cdot \big(\Rot \cdot \Ysig\big) + \Xemzz \frac{1}{\omega \mu_0} \AlphaT \cdot \big(\Rot\cdot \kpar\big) \Big] \Bigg\} \cdot \vEi.
\end{aligned}
\end{equation}
\end{subequations}

\section{Kerker effect}
The transmission and reflection coefficients for an isotropic metasurface with the susceptibilities $\chiee{xx}$ and $\chimm{yy}$ that is embedded in an homogeneous background medium are given by~\cite{achouri2021}

\begin{subequations}
\begin{align*}
    R &= \frac{2jk \left ( \chiee{xx} - \chimm{yy}\right )}{\left ( 2 + jk\chiee{xx}\right ) \left ( 2 + jk\chimm{yy}\right )}, \\
    T &= \frac{4 + \chiee{xx}\chimm{yy}k^2}{\left ( 2 + jk\chiee{xx}\right ) \left ( 2 + jk\chimm{yy}\right )}.
\end{align*}
\end{subequations}

The Kerker condition imposes that the electric and magnetic responses cancel out each other, reducing the reflection to zero when $\chiee{xx}=\chimm{yy}$.

\begin{subequations}
\begin{align*}
    \left. R \right |_{\chiee{xx}=\chimm{yy}} &=0, \\
    \left. T \right |_{\chiee{xx}=\chimm{yy}} &= \frac{2j + \chiee{xx}k}{\left ( 2j - \chiee{xx}k\right )}.
\end{align*}
\end{subequations}

The transmission amplitude and phase are therefore $|T|=1$ and $\angle T = \arctan \left ( \frac{4k\chiee{xx}}{k^2\chiee{xx 2}-4}\right )$. The phase is zero if $\chiee{xx}=0$, signifying that invisibility is achievable if there is no metasurface, which is the trivial solution. Another solution would consists in implementing scattering particles exhibiting double Lorentzian resonances, which would approach $\chiee{xx}\approx 0$ at a frequency somewhere in between the two resonances.

\section{\textit{z}-symmetry breaking with substrate}
In this section, we prove that the asymmetry provided by two different background media may be modeled as a \textit{z}-asymmetric unit cell. This configuration is formally equivalent to a metasurface exhibiting an effective bianisotropic response induced by the substrate.

Let us consider only three susceptibilities, such as $\chiee{xx\prime}, \chimm{yy\prime}$ and $\chiem{xy\prime}$. We consider the metasurface to be reciprocal and is surround by vacuum. This metasurface then exhibits the scattering coefficients in terms of the mentioned susceptibilities,
\begin{subequations} \label{bianisotropic_coefficients}
\begin{align}
t_{s,12} &=
\frac{
\chiee{xx\prime}\chimm{yy\prime}k^{2}
-\left(2j+\chiem{xy\prime}\,k\right)\left(2j-\chiem{xy\prime}\,k\right)
}{
2jk\left(\chiee{xx\prime}+\chimm{yy\prime}\right)
-\left(\chiem{xy\prime}\right)^{2}k^{2}
+4-\chiee{xx\prime}\chimm{yy\prime}k^{2}
},
\\[6pt]
r_{s,12} &=
\frac{
2jk\left(\chimm{yy\prime}-\chiee{xx\prime}-2\chiem{xy\prime}\right)
}{
2jk\left(\chiee{xx\prime}+\chimm{yy\prime}\right)
-\left(\chiem{xy\prime}\right)^{2}k^{2}
+4-\chiee{xx\prime}\chimm{yy\prime}k^{2}
},
\\[6pt]
r_{s,21} &=
\frac{
2jk\left(\chimm{yy\prime}-\chiee{xx\prime}+2\chiem{xy\prime}\right)
}{
2jk\left(\chiee{xx\prime}+\chimm{yy\prime}\right)
-\left(\chiem{xy\prime}\right)^{2}k^{2}
+4-\chiee{xx\prime}\chimm{yy\prime}k^{2}
},
\\[6pt]
t_{s,21} &= t_{s,12}.
\end{align}
\end{subequations}
Now that we have the scattering coefficients of a reciprocal bianisotropic sheet, we consider a more conventional implementation where the metasurface is intrinsically \emph{anisotropic} (no magnetoelectric coupling), but is placed on top of a substrate (or separated from an interface by a spacer). We show that this \textit{extrinsic} $z$-asymmetry generates an \emph{effective} omega-type bianisotropy.
\begin{figure}[H]
    \centering
    \includegraphics[width=0.6\linewidth]{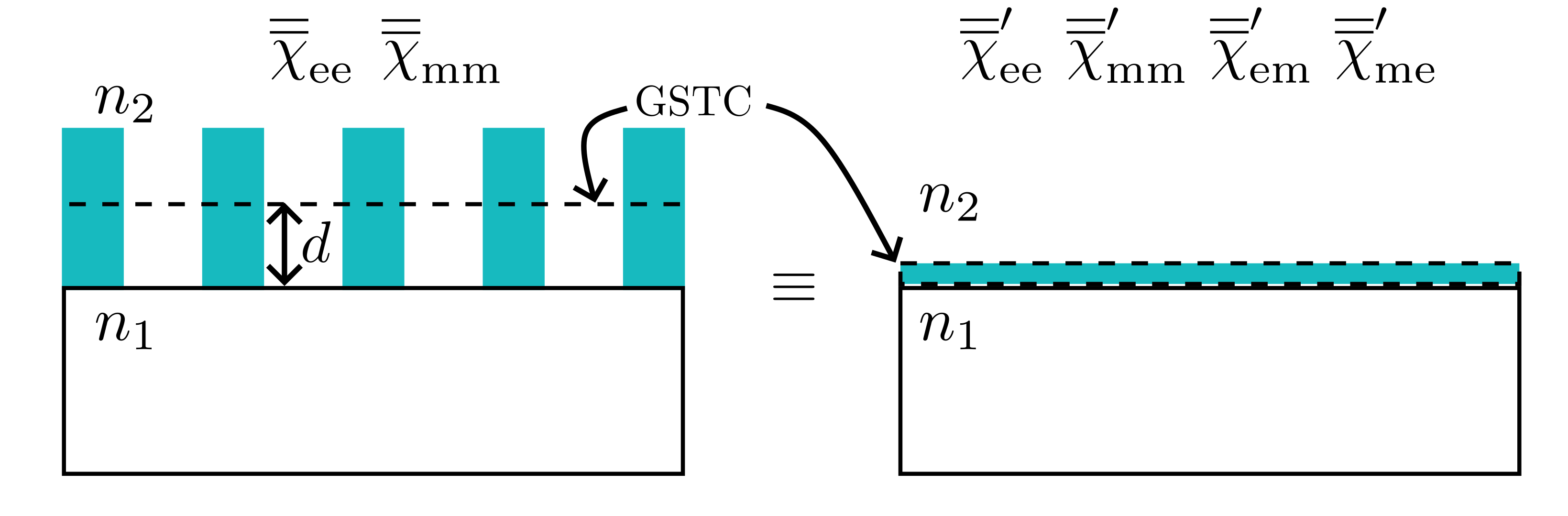}
    \caption{A metasurface exhibiting only anisotropic susceptibility as an equivalent of a metasurface that exhibits bianisotropic responses caused by dissimilar surrounding media.}
    \label{ms_equivalence}
\end{figure}

\subsection{Anisotropic sheet displaced from an interface}
We consider an anisotropic metasurface embedded in medium~2 and located at a distance $d$ from the planar interface separating medium~1 and medium~2, as depicted in Fig.~\ref{ms_equivalence}. The metasurface is assumed reciprocal and purely anisotropic,
\begin{equation}
\Xee \neq 0,\qquad \Xmm \neq 0, \qquad \Xem=\Xme=0,
\end{equation}
so that any magnetoelectric coupling appearing in the overall response must originate from the environment (substrate/spacer).

A convenient way to compute the overall scattering is to use transfer matrix method. Denoting by $\mathbf{M}_1$ and $\mathbf{M}_2$ the interface matrices (from medium~1 to medium~2 and back to the reference plane) and by $\mathbf{M}_d$ the propagation matrix of the spacer of thickness $d$, the total transfer matrix reads
\begin{equation}
\mathbf{M}=\mathbf{M}_2\,\mathbf{M}_s\,\mathbf{M}_d\,\mathbf{M}_1,
\end{equation}
where $\mathbf{M}_s$ is the transfer matrix of the \emph{intrinsically anisotropic} metasurface (obtained from the same relations as above by setting $\chiem{xy}=0$), and
\begin{equation}
\mathbf{M}_d=
\begin{pmatrix}
e^{-j\phi} & 0\\
0 & e^{+j\phi}
\end{pmatrix},
\end{equation}
with $\phi=n_2kd$ the added phase in the spacer.
Writing $\mathbf{M}=\begin{pmatrix}A & B\\ C & D\end{pmatrix}$, the overall scattering coefficients follow from the standard relations
\begin{equation}
t_{12}=\frac{AD-BC}{D},\qquad r_{12}=-\frac{C}{D},\qquad r_{21}=\frac{B}{D}.
\end{equation}
from which we deduce the scattering coefficients,
\begin{subequations} \label{aniso_coefficients}
\begin{align}
t_{12} &=
\frac{
2 e^{i d k n_2} \, n_1 n_2
\left(4 + k^2 \chiee{xx} \chimm{yy}\right)
}{
e^{2 i d k n_2}(n_1+n_2)
\left(2 n_2 + i k \chiee{xx}\right)
\left(2 + i k n_2 \chimm{yy}\right)
+
2 i k (n_1-n_2)
\left(-\chiee{xx} + n_2^2 \chimm{yy}\right),
}
\\[10pt]
r_{12} &=
\frac{
- e^{2 i d k n_2}( -n_1+n_2 )
\left(2 n_2 + i k \chiee{xx}\right)
\left(-2 i + k n_2 \chimm{yy}\right)
+
2 k (n_1+n_2)
\left(-\chiee{xx} + n_2^2 \chimm{yy}\right)
}{
e^{2 i d k n_2}(n_1+n_2)
\left(2 n_2 + i k \chiee{xx}\right)
\left(-2 i + k n_2 \chimm{yy}\right)
+
2 k (n_1-n_2)
\left(-\chiee{xx} + n_2^2 \chimm{yy}\right),
}
\\[10pt]
r_{21} &=
\frac{
(n_1-n_2)
\left(2 n_2 - i k \chiee{xx}\right)
\left(2 i + k n_2 \chimm{yy}\right)
+
2 e^{2 i d k n_2} k (n_1+n_2)
\left(-\chiee{xx} + n_2^2 \chimm{yy}\right)
}{
e^{2 i d k n_2}(n_1+n_2)
\left(2 n_2 + i k \chiee{xx}\right)
\left(-2 i + k n_2 \chimm{yy}\right)
+
2 k (n_1-n_2)
\left(-\chiee{xx} + n_2^2 \chimm{yy}\right)
}.
\end{align}
\end{subequations}
\subsection{Equivalent bianisotropic sheet}

We now seek an \emph{equivalent} reciprocal bianisotropic sheet located at the interface between the two media, characterized by the equivalent susceptibilities
$\chiee{xx^\prime}$, $\chimm{yy^\prime}$ and $\chiem{xy^\prime}$, whose scattering coefficients are given by Eqs.~\eqref{bianisotropic_coefficients}.
By enforcing
\begin{equation}
t_{12}=t_{s,12},\qquad r_{12}=r_{s,12},\qquad r_{21}=r_{s,21},
\end{equation}
one obtains closed-form expressions for the effective parameters. In the particularly relevant case where the metasurface is embedded in the same medium as the spacer (medium~2) and only the substrate breaks the $z$-symmetry, the effective susceptibilities can be written as
\begin{subequations}\label{effective_chi_substrate}
\begin{align}
\chiee{xx^\prime} &=
\frac{
4 k\,\chiee{xx}\cos(kd)+\Big(4-k^{2}\chiee{xx}\chimm{yy}\Big)\sin(kd)
}{
k\Big[-2\cos\!\Big(\tfrac{kd}{2}\Big)+k\,\chiee{xx}\sin\!\Big(\tfrac{kd}{2}\Big)\Big]
 \Big[-2\cos\!\Big(\tfrac{kd}{2}\Big)+k\,\chimm{yy}\sin\!\Big(\tfrac{kd}{2}\Big)\Big]
},
\\[4pt]
\chimm{yy^\prime} &=
\frac{
4 k\,\chimm{yy}\cos(kd)+\Big(4-k^{2}\chiee{xx}\chimm{yy}\Big)\sin(kd)
}{
k\Big[-2\cos\!\Big(\tfrac{kd}{2}\Big)+k\,\chiee{xx}\sin\!\Big(\tfrac{kd}{2}\Big)\Big]
 \Big[-2\cos\!\Big(\tfrac{kd}{2}\Big)+k\,\chimm{yy}\sin\!\Big(\tfrac{kd}{2}\Big)\Big]
},
\\[4pt]
\chiem{xy^\prime} &=
-\frac{
4j\,\big(\chiee{xx}-\chimm{yy}\big)\,\sin(kd)
}{
4+k^{2}\chiee{xx}\chimm{yy}
+\big(4-k^{2}\chiee{xx}\chimm{yy}\big)\cos(kd)
-2k\big(\chiee{xx}+\chimm{yy}\big)\sin(kd)
}.
\end{align}
\end{subequations}

Equations~\eqref{effective_chi_substrate} explicitly shows that an \emph{extrinsic} $z$-asymmetry (here introduced by the substrate/spacer) generates a nonzero effective magnetoelectric coupling $\chiem{xy^\prime}$ even though the original metasurface satisfies $\chiem{xy}=0$. 

% In particular, $\chiem{xy^\prime}$ vanishes in the following important cases:
% \begin{itemize}
% \item \textbf{Balanced electric/magnetic (Kerker) response:} $\chiee{xx}=\chimm{yy}$, for which the metasurface behaves as a symmetric Huygens-type sheet and no electromagnetic coupling is induced.
% \item \textbf{No phase delay:} $kd=m\pi$ ($m\in\mathbb{Z}$), where $\sin(kd)=0$.
% \end{itemize}

These results prove that an asymmetric electromagnetic environment (dissimilar media or a substrate-backed implementation) is formally equivalent to a $z$-asymmetric unit cell exhibiting effective bianisotropy, thereby providing a practical route to realize the magnetoelectric terms required by the generalized invisibility conditions derived in the previous section.
\subsection{Cancellation of the anisotropic terms}
\label{app:cancel_aniso}

The effective parameters of the equivalent bianisotropic sheet are given by
Eqs.~\eqref{effective_chi_substrate}. In particular, the effective anisotropic
susceptibilities read
\begin{align}
\chiee{xx^\prime} &=
\frac{
4 k\,\chiee{xx}\cos(kd)+\Big(4-k^{2}\chiee{xx}\chimm{yy}\Big)\sin(kd)
}{
k\Big[-2\cos\!\Big(\tfrac{kd}{2}\Big)+k\,\chiee{xx}\sin\!\Big(\tfrac{kd}{2}\Big)\Big]
 \Big[-2\cos\!\Big(\tfrac{kd}{2}\Big)+k\,\chimm{yy}\sin\!\Big(\tfrac{kd}{2}\Big)\Big]
},
\label{eq:chiee_prime_again}
\\[4pt]
\chimm{yy^\prime} &=
\frac{
4 k\,\chimm{yy}\cos(kd)+\Big(4-k^{2}\chiee{xx}\chimm{yy}\Big)\sin(kd)
}{
k\Big[-2\cos\!\Big(\tfrac{kd}{2}\Big)+k\,\chiee{xx}\sin\!\Big(\tfrac{kd}{2}\Big)\Big]
 \Big[-2\cos\!\Big(\tfrac{kd}{2}\Big)+k\,\chimm{yy}\sin\!\Big(\tfrac{kd}{2}\Big)\Big]
}.
\label{eq:chimm_prime_again}
\end{align}
To enforce \(\chiee{xx^\prime}=\chimm{yy^\prime}=0\) with a finite response, the
numerators of \eqref{eq:chiee_prime_again}--\eqref{eq:chimm_prime_again} must vanish
while the denominator remains nonzero. This yields the coupled conditions
\begin{subequations}\label{eq:num_zero_conditions}
\begin{align}
4k\,\chiee{xx}\cos(kd)+\big(4-k^2\chiee{xx}\chimm{yy}\big)\sin(kd) &= 0,
\label{eq:num_zero_a}\\
4k\,\chimm{yy}\cos(kd)+\big(4-k^2\chiee{xx}\chimm{yy}\big)\sin(kd) &= 0.
\label{eq:num_zero_b}
\end{align}
\end{subequations}
Subtracting \eqref{eq:num_zero_b} from \eqref{eq:num_zero_a} gives
\begin{equation}
4k\big(\chiee{xx}-\chimm{yy}\big)\cos(kd)=0,
\end{equation}
which implies two solution branches.

\paragraph*{Branch~1: balanced sheet \(\chiee{xx}=\chimm{yy}\).}
If \(\chiee{xx}=\chimm{yy}\), Eqs.~\eqref{eq:num_zero_conditions} reduce to a single
constraint
\begin{equation}
4k\,\chiee{xx}\cos(kd)+\big(4-k^2\chiee{xx}{}^2\big)\sin(kd)=0,
\end{equation}
which fixes \(\chiee{xx}\) (and hence \(\chimm{yy}\)) for a chosen electrical thickness \(kd\).
In this branch, the induced omega coupling \(\chiem{xy^\prime}\) also vanishes
because it is proportional to \(\chiee{xx}-\chimm{yy}\) [see Eq.~\eqref{effective_chi_substrate}],
so the equivalent sheet remains purely anisotropic.

\paragraph*{Branch~2: quarter-wave spacer \(\cos(kd)=0\).}
A more relevant branch for generating effective bianisotropy is obtained by choosing
\begin{equation}
\cos(kd)=0
\quad\Longleftrightarrow\quad
kd=\frac{\pi}{2}+m\pi,\qquad m\in\mathbb{Z},
\label{eq:quarter_wave_condition}
\end{equation}
for which \(\sin(kd)=\pm 1\). Then Eqs.~\eqref{eq:num_zero_conditions} become
\begin{equation}
\big(4-k^2\chiee{xx}\chimm{yy}\big)\sin(kd)=0,
\end{equation}
and hence the simultaneous cancellation of the effective anisotropic terms is achieved by the
\emph{self-dual} condition
\begin{equation}
k^2\chiee{xx}\chimm{yy}=4
\qquad\Longleftrightarrow\qquad
\chimm{yy}=\frac{4}{k^2\,\chiee{xx}}.
\label{eq:self_dual_app}
\end{equation}
Under \eqref{eq:quarter_wave_condition}--\eqref{eq:self_dual_app}, one has
\(\chiee{xx^\prime}=\chimm{yy^\prime}=0\) (provided the denominator of
\eqref{eq:chiee_prime_again}--\eqref{eq:chimm_prime_again} is nonzero), while the induced omega coupling remains,
in general, nonzero.

Using Eq.~\eqref{effective_chi_substrate}, the effective magnetoelectric coupling becomes
\begin{equation}
\chiem{xy\prime}
=
-\frac{4j\,\big(\chiee{xx}-\chimm{yy}\big)\,\sin(kd)}
{8-2k\big(\chiee{xx}+\chimm{yy}\big)\sin(kd)}
\qquad\text{for}\qquad
\cos(kd)=0,\;\; k^2\chiee{xx}\chimm{yy}=4,
\end{equation}
showing explicitly that the substrate/spacer converts an intrinsically anisotropic sheet into an
\(\Omega\)-type bianisotropic response. In particular, choosing \(kd=\pi/2\) (mod \(\pi\)) yields a purely
imaginary \(\chiem{xy^\prime}\) for lossless \(\chiee{xx},\chimm{yy}\), consistent with passive omega coupling
under the adopted time convention.
\newpage
\section{Unit Cell Simulations}
\begin{figure}[H]
    \centering
    \includegraphics[width=0.7\linewidth]{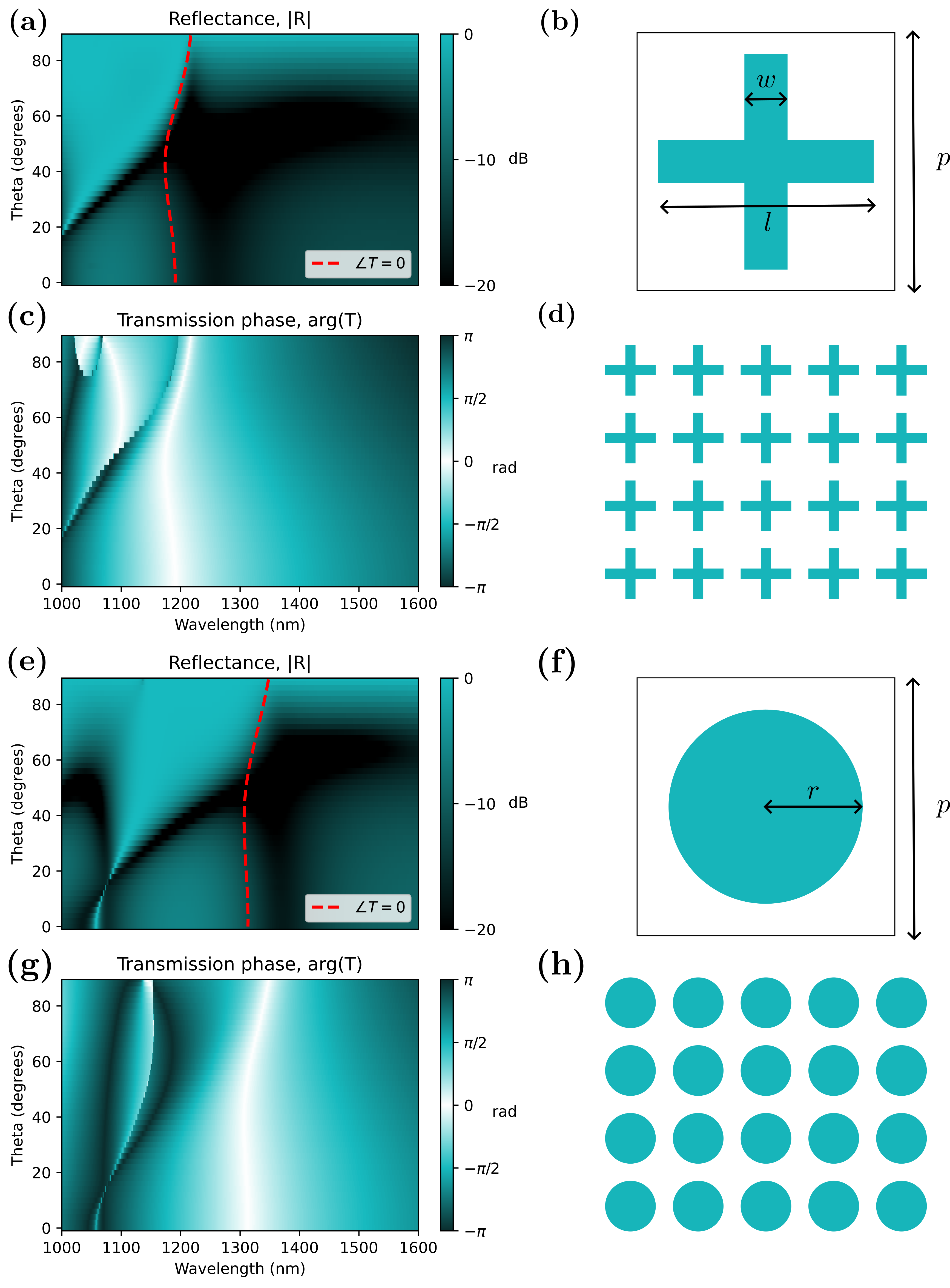}
    \caption{Invisibility simulation. \textbf{(a)},\textbf{(e)} Simulation of the reflectance $|R|$ and the transmission phase $\angle T$. The dashed line on the top figure shows the zero phase of the transmission which crosses the reflectance below -20 dB. b) The geometric parameters of the cross unit cell is : $p$ = 400nm, $l$ = 300nm, $w$ = 100nm, $t$ = 630nm. c),f) The metasurface is placed on top of a substrate of SiO$_2$ and the cross and cylinders are made of Si. d) The geometric parameters of the cylinder unit cell is : $p$ = 400nm, $r$ = 150nm, $t$ = 600nm.}
    \label{fig_si_sim_cross}
\end{figure}

\end{document}